\documentclass[twocolumn]{aastex631}

\usepackage{amsmath}

\newcommand{\psrzero}{PSR~J0058$-$7218}
\newcommand{\psrfive}{PSR~J0537$-$6910}
\newcommand{\psrone}{PSR~J1101$-$6101}
\newcommand{\psrfour}{PSR~J1412+7922}
\newcommand{\psreightone}{PSR~J1813$-$1749}
\newcommand{\psreight}{PSR~J1849$-$0001}
\newcommand{\Pdot}{\dot{P}}
\newcommand{\nudot}{\dot{\nu}}
\newcommand{\nuddot}{\ddot{\nu}}
\newcommand{\tauc}{\tau_{\rm c}}
\newcommand{\Edot}{\dot{E}}
\newcommand{\nig}{n_{\rm ig}}
\newcommand{\Ag}{A_{\rm g}}
\newcommand{\tobs}{T_{\rm obs}}
\newcommand{\nudotg}{\dot{\nu}_{\rm g}}
\newcommand{\nudotig}{\dot{\nu}_{\rm ig}}
\newcommand{\nuddotig}{\ddot{\nu}_{\rm ig}}
\newcommand{\Ng}{N_{\rm g}}
\newcommand{\tg}{T_{\rm g}}

\shorttitle{Timing X-ray pulsars, including \psrzero\ and \psrfive}
\shortauthors{Ho et al.}
\graphicspath{{./}{figures/}}

\begin{document}

\title{Timing six energetic rotation-powered X-ray pulsars, including the \\ fast-spinning young \psrzero\ and Big Glitcher \psrfive}

\correspondingauthor{Wynn C. G. Ho}
\email{wynnho@slac.stanford.edu}

\author[0000-0002-6089-6836]{Wynn C. G. Ho}
\affiliation{Department of Physics and Astronomy,
Haverford College, 370 Lancaster Avenue, Haverford, PA 19041, USA}
\author{Lucien Kuiper}
\affiliation{SRON-Netherlands Institute for Space Research,
Niels Bohrweg 4, 2333 CA, Leiden, Netherlands}
\author{Crist\'obal M. Espinoza}
\affiliation{Departamento de F\'isica, Universidad de Santiago de Chile (USACH),
Av. Victor Jara 3493, Estaci\'on Central, Chile}
\affiliation{Center for Interdisciplinary Research in Astrophysics and
Space Sciences (CIRAS), Universidad de Santiago de Chile, Santiago, Chile}
\author[0000-0002-6449-106X]{Sebastien Guillot}
\affiliation{IRAP, CNRS, 9 avenue du Colonel Roche, BP 44346,
31028 Toulouse Cedex 4, France}
\affiliation{Universit\'e de Toulouse, CNES, UPS-OMP, 31028 Toulouse, France}
\author{Paul S. Ray}
\affiliation{Space Science Division, U.S. Naval Research Laboratory,
Washington, DC, 20735, USA}
\author[0000-0002-7833-0275]{D. A. Smith}
\affiliation{Laboratoire d'Astrophysique de Bordeaux, CNRS and Universit\'e
Bordeaux, B18N, all\'ee Geoffroy Saint-Hilaire, 33615 Pessac, France}
\author{Slavko Bogdanov}
\affiliation{Columbia Astrophysics Laboratory,
Columbia University, 550 West 120th Street, New York, NY 10027, USA}
\author{Danai Antonopoulou}
\affiliation{Jodrell Bank Centre for Astrophysics, School of Physics
and Astronomy, University of Manchester, Manchester M13 9PL, UK}
\author{Zaven Arzoumanian}
\affiliation{X-Ray Astrophysics Laboratory,
NASA Goddard Space Flight Center, Greenbelt, MD, 20771, USA}
\author{Micha{\l} Bejger}
\affiliation{INFN Sezione di Ferrara, Via Saragat 1, 44122 Ferrara, Italy}
\affiliation{Nicolaus Copernicus Astronomical Center,
Polish Academy of Sciences, Bartycka 18, 00-716 Warsaw, Poland}
\author{Teruaki Enoto}
\affiliation{Extreme Natural Phenomena RIKEN Hakubi Research Team,
Cluster for Pioneering Research, RIKEN, Wako 351-0198, Japan}
\author{Paolo Esposito}
\affiliation{Scuola Universitaria Superiore IUSS Pavia,
Palazzo del Broletto, piazza della Vittoria 15, 27100 Pavia, Italy}
\affiliation{Istituto Nazionale di Astrofisica,
IASF--Milano, via Alfonso Corti 12, 20133 Milano, Italy}
\author{Alice K. Harding}
\affiliation{Theoretical Division, Los Alamos National Laboratory,
Los Alamos, NM 87545, USA}
\author{Brynmor Haskell}
\affiliation{Nicolaus Copernicus Astronomical Center,
Polish Academy of Sciences, Bartycka 18, 00-716 Warsaw, Poland}
\author[0000-0003-0771-6581]{Natalia Lewandowska}
\affiliation{Swarthmore College, 500 College Avenue, Swarthmore, PA 19081, USA}
\author{Chandreyee Maitra}
\affiliation{Max-Planck-Institut f\"ur extraterrestrische Physik,
Gie\ss enbachstra\ss e 1, 85748 Garching, Germany}
\author{Georgios Vasilopoulos}
\affiliation{Universit\'e de Strasbourg, CNRS,
Observatoire astronomique de Strasbourg, UMR 7550, F-67000 Strasbourg, France}

\begin{abstract}
Measuring a pulsar's rotational evolution is crucial to understanding
the nature of the pulsar.
Here we provide updated timing models for the rotational evolution of
six pulsars,
five of which are rotation phase-connected using primarily NICER X-ray data.
For the newly-discovered fast energetic young pulsar, \psrzero,
we increase the baseline of its timing model from 1.4 days to 8~months
and not only measure more precisely its spin-down rate
$\nudot=(-6.2324\pm0.0001)\times10^{-11}\mbox{ Hz s$^{-1}$}$ but also
for the first time the second time derivative of spin rate
$\nuddot=(4.2\pm0.2)\times10^{-21}\mbox{ Hz s$^{-2}$}$.
For the fastest and most energetic young pulsar, \psrfive\
(with 16~ms spin period),
we detect 4 more glitches, for a total of 15 glitches over 4.5 years of
NICER monitoring, and show that its spin-down behavior continues to set this
pulsar apart from all others, including a long-term braking index
$n=-1.234\pm0.009$ and interglitch braking indices that asymptote
to $\lesssim 7$ for long times after a glitch.
For \psrone, we measure a much more accurate spin-down rate that
agrees with a previous value measured without phase-connection.
For \psrfour\ (also known as Calvera), we extend the baseline of its
timing model from our previous 1-year model to 4.4 years,
and for \psreight, we extend the baseline from 1.5 years to 4.7 years.
We also present a long-term timing model of the energetic pulsar,
\psreightone, by fitting previous radio and X-ray spin frequencies from
2009--2019 and new ones measured here using 2018 NuSTAR and 2021 Chandra data.
\end{abstract}

\section{Introduction} \label{sec:intro}

\begin{figure}
\plotone{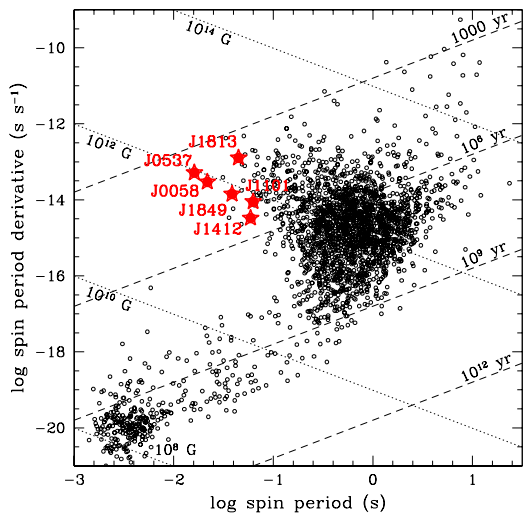}
\caption{Pulsar spin period $P$ and spin period time derivative $\Pdot$.
Circles denote pulsars whose values are taken from the ATNF Pulsar Catalogue
(\citealt{manchesteretal05}, version~1.66), and stars indicate pulsars
considered in this work
(see Table~\ref{tab:psr}).
Dashed lines indicate characteristic age $\tauc\equiv P/2\Pdot$, and
dotted lines indicate
magnetic field strength $B=3.2\times10^{19}\mbox{ G }(P\Pdot)^{1/2}$.
\label{fig:ppdot}}
\end{figure}

Accurate measurements of the evolution of a pulsar's rotation rate
are vital to inferring properties of and classifying the pulsar.
For example, the spin period $P$ ($=1/\nu$, where $\nu$ is spin frequency)
and spin period time
derivative $\Pdot$ enable estimates of a pulsar's age
(via the characteristic age $\tauc\equiv P/2\Pdot$)
and magnetic field strength $B$
[$\approx3.2\times10^{19}\mbox{ G }(P\Pdot)^{1/2}$]
and give the rate of rotational energy loss
$\Edot\approx4.0\times10^{46}\mbox{ erg s$^{-1}$}\Pdot/P^3$
(e.g., \citealt{shapiroteukolsky83}).
For neutron star classification, Figure~\ref{fig:ppdot} demonstrates
that the $\sim 3000$ known pulsars reside in distinct regions in the
$P$--$\Pdot$ parameter space.

Regular monitoring observations of pulsars enable detection of glitches,
which are sudden changes in the spin frequency of a pulsar
that mostly occur in pulsars with $\tauc<10^7\mbox{ yr}$
\citep{espinozaetal11,yuetal13,liuetal21,loweretal21,basuetal22},
as well as timing irregularities (i.e., timing noise).
Most glitches are thought to be due to unpinning of superfluid
vortices in the neutron star \citep{andersonitoh75} and
can be used to infer properties of the superfluid interior
\citep{linketal99,lyneetal00,anderssonetal12,chamel13,hoetal15}.

\begin{deluxetable*}{ccccccccc}
\tablecaption{Properties of pulsars considered in present work\label{tab:psr}}
\tablewidth{0pt}
\tablehead{
\colhead{Pulsar} & \colhead{$P$} & \colhead{$\Pdot$} & \colhead{$\Edot$} & \colhead{$B$} & \colhead{$\tauc$} & \colhead{SNR} & \colhead{SNR age} & \colhead{$d$} \\
& (ms) & ($10^{-14}\mbox{ s s$^{-1}$}$) & ($10^{37}\mbox{ erg s$^{-1}$}$) & ($10^{12}\mbox{ G}$) & (kyr) & & (kyr) & (kpc)
}
\startdata
\psrzero & 21.8 & 2.95 & 11 & 0.81 & 11.7 & IKT~16 & 14.7 [1] & 62 [2] \\
\psrfive & 16.2 & 5.21 & 49 & 0.93 & 4.91 & N157B & 1--5 [3] & 49.6 [4] \\
\psrone  & 62.8 & 0.893 & 0.14 & 0.76 & 111 & MSH 11$-$61A & 10--30 [5] & $7\pm1$ [6] \\
\psrfour & 59.2 & 0.330 & 0.064 & 0.45 & 285 & \nodata & \nodata & $\lesssim$3.3 [7] \\
\psreightone & 44.7 & 12.7 & 5.7 & 2.4 & 5.58 & G12.82$-$0.02 & $<3$ [8] & 6--14 [9] \\
\psreight & 38.5 & 1.42 & 0.99 & 0.75 & 43.1 & \nodata & \nodata & 7 [10] \\
\enddata
\tablecomments{Spin period $P$ and spin period time derivative $\Pdot$,
spin-down luminosity $\Edot=4.0\times10^{46}\mbox{ erg s$^{-1}$ }\Pdot/P^3$,
magnetic field $B=3.2\times10^{19}\mbox{ G }(P\Pdot)^{1/2}$,
characteristic age $\tauc\equiv P/2\Pdot$,
supernova remnant (SNR) association and age, and distance $d$.}
\tablerefs{[1]: \citet{owenetal11}, [2]: \citet{graczyketal20},
[3]: \cite{chenetal06}, [4]: \citet{pietrzynskietal19},
[5]: \cite{garciaetal12}, [6]: \citet{reynosoetal06},
[7]: \citet{mereghettietal21},
[8]: \cite{broganetal05}, [9]: \citet{camiloetal21},
[10]: \citet{gotthelfetal11}.}
\vspace{-2em}
\end{deluxetable*}

Pulsars are also potential sources of detectable gravitational waves that
are continuously emitted for the lifetime of a pulsar and occur at
gravitational wave frequencies proportional to the pulsar spin frequency.
The most sensitive searches for continuous gravitational waves are
those targeting known pulsars that have an accurate contemporaneous
rotation phase-connected timing solution since such a model greatly reduces
the parameter space of searches \citep{riles17,sieniawskabejger19}.
Searches of data from the most recent 2019--2020 LIGO/Virgo observing run (O3)
include five of the six pulsars studied here (see Table~\ref{tab:psr}),
with the exception being
\psrzero\ since its spin properties only became known in 2021.
Constraints are placed on the size of a gravitational wave-producing
mountain in each pulsar \citep{abbottetal21,abbottetal22,abbottetal22b}
and the size of an r-mode fluid oscillation in \psrfive\ \citep{abbottetal21b}.
A search is also made for gravitational waves produced by glitches
of \psrfive\ and \psreightone\ \citep{abbottetal22b}.
Maintaining X-ray monitoring of the pulsars studied here is necessary
to enable stronger constraints, and even detection, in more sensitive
data that will be collected in future gravitational wave observing runs,
such as the next one (O4) scheduled to begin in 2023 March.

Here we report on timing models for six pulsars
(see Table~\ref{tab:psr}), which are the result of monitoring data from
Chandra \citep{weisskopfetal02}, NICER \citep{gendreauetal16},
and NuSTAR \citep{harrisonetal13}.
The spin pulsation of each pulsar is
only detectable at X-ray energies,
except the radio pulsation detection recently reported for \psreightone\
(\citealt{camiloetal21}; see below).
Our results include new long-term phase-connected timing models
for \psrzero\ and \psrone\ covering time baselines of 8~months and
1.7~years, respectively.
The baselines of the phase-connected timing models of \psrfive,
\psrfour, and \psreight\ are extended up to three to four times those
previously reported, and four new glitches of \psrfive\ are presented.
For \psreightone, we fit spin frequency measurements made over
12~years by Chandra, Green Bank, NICER, NuSTAR, and XMM-Newton,
including two new measurements by Chandra in 2021, to obtain an
updated spin-down rate, assuming a constant linear decline.
We also report on pulsation searches of Fermi Gamma-Ray Burst Monitor
(GBM; \citealt{meeganetal09}) and Large Area Telescope
(LAT; \citealt{atwoodetal09}) data using our
timing models of \psrzero, \psrone, \psrfour, and \psreight.

An outline of the paper is as follows.
Section~\ref{sec:psr} briefly summarizes relevant information for
each of our six pulsars.
Section~\ref{sec:data} describes the data and its processing.
Section~\ref{sec:results} presents our results,
including timing models for each of the six pulsars.
Section~\ref{sec:discuss} summarizes and discusses some implications
of the work presented here.

\section{Summary of pulsars} \label{sec:psr}

\psrzero\ is a newly identified fast spinning young pulsar
in the 14.7~kyr supernova remnant IKT~16 in the Small Magellanic Cloud
and associated with a pulsar wind nebula \citep{owenetal11,maitraetal15}.
\citet{maitraetal21} measured for the first time the timing properties
of \psrzero\ using
XMM-Newton EPIC-pn in small window mode with a time resolution of 5.7~ms,
including $\nudot=(-6.1\pm0.6)\times 10^{-11}\mbox{ Hz s$^{-1}$}$
(1$\sigma$ error) within the 118~ks exposure.
The pulsar has a narrow single peak pulse profile and high pulsed fraction
of $\approx 70\%$ in the 0.4--10~keV band.
Its rapid spin-down rate means \psrzero\ has the fourth highest measured
spin-down luminosity $\Edot$ among the more than 3000 known pulsars.

\psrfive\ is the fastest spinning young pulsar and is in the
1--5~kyr supernova remnant N157B in the Large Magellanic Cloud
\citep{wanggotthelf98,chenetal06}.
The pulsar has a narrow single peak pulse profile and pulsed fraction
of $\sim 20\%$ in the 2--8~keV band \citep{marshalletal98,kuiperhermsen15}.
While its spin frequency decreases over the more than 17 years
of combined observations (1998--2011 with RXTE and 2017--2022 with NICER),
a remarkable 60 glitches are measured thus far, including 15 by NICER.
This yields an average glitch rate of $\sim 3.5\mbox{ yr$^{-1}$}$
and glitch magnitudes larger than those seen in most pulsars
\citep{middleditchetal06,antonopoulouetal18,ferdmanetal18,hoetal20,abbottetal21};
thus \psrfive\ is known as the Big Glitcher \citep{marshalletal04}.
Its glitches are unusual in their predictability, in particular
there is a correlation seen between the size of its glitches
$\Delta\nu$ and time to its next glitch
\citep{middleditchetal06,antonopoulouetal18,ferdmanetal18,hoetal20}.
Its timing properties are also unusual, with a braking index
$n\equiv\nuddot\nu/\nudot^2=-1.25\pm0.01$ (1$\sigma$ error)
over the long-term (17-year duration) and
a value that approaches $\lesssim 7$ over the short-term
($\sim 100\mbox{ day}$) between glitches.

\psrone\ is associated with the hard X-ray source
IGR~J11014$-$6103 (Lighthouse nebula) and
supernova remnant MSH~11$-$61A (also known as G290.1$-$0.8).
The remnant has an age of 10--30~kyr \citep{garciaetal12},
which implies \psrone\ is a fast-moving pulsar \citep{tomsicketal12}
with a velocity of 800--2400~km~s$^{-1}$ that is consistent with
its weak proper motion upper limit \citep{pavanetal16}.
The pulsar has a broad single peak pulse profile, pulsed fraction of
$\approx 50\%$ in the 0.5--10~keV band, and spin-down rate of
$\nudot=(-2.17\pm0.13)\times10^{-12}\mbox{ Hz s$^{-1}$}$ (1$\sigma$ error)
that comes from two spin frequency measurements
using XMM-Newton EPIC-pn in small window mode
separated by about one year in 2013 and 2014 \citep{halpernetal14}.
\citet{kuiperhermsen15} measure a pulsed fraction of $\sim64\%$ in
the 2--10~keV band using the 2013 data.

\psrfour\ (also known as Calvera) is a high Galactic latitude pulsar
whose timing parameters suggest the pulsar is relatively middle-aged
and energetic \citep{zaneetal11,halpernetal13,halperngotthelf15}.
The pulsar could be a descendant of a member of the central compact object
(CCO) class of neutron stars (see \citealt{deluca17}, for review)
whose supernova remnant is no longer visible.
Alternatively, \psrfour\ may simply be a normal rotation-powered pulsar
\citep{mereghettietal21}.
A major contributor to the uncertainty in its classification is its
distance, with a lower limit of about 200~pc and a value up to 3.8~kpc
from spectral fitting \citep{halpernetal13,halperngotthelf15,mereghettietal21}.
Monitoring during the first year of NICER in 2017--2018 yielded
a phase-connected timing model \citep{bogdanovetal19}, and
subsequent work extended the timing model to over three years
using NICER data through 2021 February \citep{mereghettietal21}.

\psreightone\ is a highly energetic pulsar that produces a
pulsar wind nebula and is associated with the gamma-ray/TeV source
IGR~J18135$-$1751/HESS~J1813$-$178.
The pulsar is located in the young ($<$3~kyr) supernova remnant
G12.82$-$0.02 \citep{broganetal05}, and its proper motion from the
center of the remnant implies an age of 1000--2200~yr \citep{dzibrodriguez21}.
\psreightone\ has a broad single peak X-ray pulse profile and
pulsed fraction of 50\% in the 2--10~keV band
(\citealt{gotthelfhalpern09,halpernetal12,kuiperhermsen15};
see also \citealt{hoetal20a}).
A phase-connected timing model spanning 37 days in 2019 was
determined using NICER data \citep{hoetal20a}.
More recently, \citet{camiloetal21} report radio observations at high
frequencies using the Green Bank Telescope in which highly scattered
radio pulses are detected for the first time for \psreightone.
They also perform a linear fit of individual spin frequencies
measured in 2009, 2011, and 2012 in X-ray and 2012 and 2015 in radio
and find a spin-down rate
$\nudot=(-6.3364\pm0.0025)\times10^{-11}\mbox{ Hz s$^{-1}$}$ (1$\sigma$ error),
which is within $3.2\sigma$ of the value determined in \citet{hoetal20a}
that includes the 2019 NICER data but not the radio data.
With confirmation of the pulsar in radio, a more reliable proper
motion is measured using the Very Large Array (VLA; \citealt{dzibrodriguez21}),
and the apparent X-ray proper motion is likely due to brightness
changes in the pulsar wind nebula, as pointed out in \citet{hoetal20a}.

\psreight\ is another young and energetic pulsar that produces a
pulsar wind nebula and is associated with the gamma-ray/TeV source
IGR~J18490$-$0000/HESS~J1849$-$000.
The pulsar has a broad single peak pulse profile and high pulsed fraction
of 77\% in the 0.06--10~keV band
(\citealt{kuiperhermsen15}; see also \citealt{bogdanovetal19}).
A phase-connected timing model spanning 20 days in 2010 was
determined using RXTE data \citep{gotthelfetal11,kuiperhermsen15},
and more recently a phase-connected timing model spanning about
1.5 years in 2017--2018 was determined using Swift and NICER data
\citep{bogdanovetal19}.

\section{Data and analysis method} \label{sec:data}

\subsection{NICER data} \label{sec:nicer}

For five of the six pulsars considered here, we use and report new
analysis of NICER data,
with the exception being Chandra and NuSTAR data of \psrone\ and \psreightone\
(see Sections~\ref{sec:nustar} and \ref{sec:chandra}).
NICER data for these five pulsars are summarized in Table~\ref{tab:data}
and are processed following a
similar procedure, which we outline and note source specific
differences below.
More details are provided in
\citet{kuiperhermsen09} for \psrzero\ and \psrone,
\citet{hoetal20} for \psrfive, and
\citet{bogdanovetal19} for \psrfour\ and \psreight.

\begin{deluxetable}{cccc}
\tablecaption{Observation log\label{tab:data}}
\tablehead{
\colhead{Telescope} & \colhead{Pulsar} & \colhead{Observation Date} & \colhead{Exposure} \\
 & & & (ks)
}
\startdata
Chandra & \psreightone & 2021 Feb 10 & 20 \\
& & 2021 Jun 23 & 20 \\
NICER & \psrzero & 2021 Jun 01--2022 Jan 25 & 163 \\
& \psrfive & 2017 Aug 17--2022 Feb 17 & 1374 \\
& \psrone & 2020 Apr 01--2021 Dec 16 & 370 \\
& \psrfour & 2017 Sep 15--2022 Feb 08 & 1379 \\
& \psreight & 2018 Feb 13--2021 Nov 16 & 260 \\
NuSTAR & \psrone & 2020 Nov 20 & 136 \\
& \psreightone & 2018 Mar 25 & 26 \\
\enddata
\end{deluxetable}

We process and filter NICER data of each pulsar using HEASoft~6.22--6.29b
\citep{heasarc14} and NICERDAS~2018-03-01\_V003--2021-08-31\_V008c.
We exclude all events from ``hot'' detector 34, which gives elevated
count rates in some circumstances, and portions of exposure accumulated
during passages through the South Atlantic Anomaly.
While NICER is sensitive to 0.25--12~keV photons,
we make an energy cut and extract only events within a specific
energy range optimized for pulsation searches.
These are 0.8--5~keV for \psrzero, 1--7~keV for \psrfive,
1.5--10~keV for \psrone, 0.37--1.97~keV for \psrfour, and
1.89--6~keV for \psreight.
For \psrzero\ and \psrone, we use cleaned standard event data which
are subsequently screened to exclude events that are collected during
high background levels as determined from light curves at 12--15~keV;
in instances when the exposure time of cleaned event data is too short
to derive an accurate time-of-arrival measurement (see below),
unfiltered event data with background filtering are used instead.
For \psrfive, \psrfour, and \psreight,
we ignore time intervals of enhanced background affecting all
detectors by constructing a light curve binned at 16~s and
removing intervals strongly contaminated by background flaring
when the count rate exceeds a threshold value.
The thresholds are $10\mbox{ c s$^{-1}$}$ for \psrfive,
$4.5\mbox{ c s$^{-1}$}$ for \psrfour, and $5\mbox{ c s$^{-1}$}$ for \psreight\
and are the same as previously used in \citet{hoetal20} for \psrfive\
and \citet{bogdanovetal19} for \psrfour\ and \psreight.
Using these filtering criteria, we obtain clean data for pulse timing analysis.

We combine sets of individual ObsIDs into merged observations, with
each merged observation yielding a single time-of-arrival (TOA) measurement.
\mbox{ObsIDs} are combined such that there is sufficient exposure to confidently
detect the spin frequency of each pulsar, with typical total exposures of
20--30~ks for \psrzero, 4--9~ks for \psrfive, 30~ks for \psrone,
6--12~ks for \psrfour, and 5--9~ks for \psreight.
Merged ObsIDs are those acquired
usually within a 3--4~day span and on rare occasions within 6--7~days.
Before performing a pulsation search, we use \texttt{barycorr} to
transform between Terrestrial Time, used for event time stamps, and
Barycentric Dynamical Time (TDB)
and to account for effects of satellite motion with respect to the barycenter.
In all timing analyses performed here unless otherwise noted
(in particular, Sections~\ref{sec:1412} and \ref{sec:1849}),
source positions (and proper motions,
if measured previously) are held fixed at the values given in corresponding
tables below, along with our adopted solar system ephemeris.

For \psrzero\ and \psrone, the search for pulsations, generation of TOAs,
and determination of timing models are conducted by following procedures
described in \citet{kuiperhermsen09}.
For \psrfive, \psrfour, and \psreight, we conduct
acceleration searches using PRESTO \citep{ransometal02},
with searches using a time bin and including a number of harmonics
that are specific to each pulsar.
In particular, these are 0.5~ms and 8 harmonics for \psrfive,
3~ms and 4 harmonics for \psrfour, and 1~ms and 4 harmonics for \psreight.
Pulsations at the spin frequency of each pulsar are usually the
strongest detected.
Data are folded at the candidate pulse frequency using \texttt{prepfold}
and a refined frequency is determined.
On occasion, further iterations are performed to obtain a more
robust measurement.
Finally, we produce a pulse profile template by fitting a set of
NICER pulse profiles with a Gaussian shape; this template is then
used to determine the TOA of each merged observation following the
unbinned maximum likelihood technique described in \citet{rayetal11}.
We use TEMPO2 \citep{hobbsetal06} to fit TOAs with a timing model
and to measure glitch parameters in the case of \psrfive.

\subsection{NuSTAR data} \label{sec:nustar}

NuSTAR observed \psrone\ on 2020 November 20 (ObsID 30601029002) for 136~ks
and \psreightone\ on 2018 March 25 (ObsID 30364003002) for 26~ks
(see Table~\ref{tab:data}).
For \psrone, we use data processed with NuSTARDAS v2.0.0 and CALDB v20200811,
and we barycenter cleaned event data extracted from
a circular region of 50\arcsec\ radius.
For \psreightone, we process data following the standard procedure with
NuSTARDAS v2.1.1 and CALDB v20210210 and use cleaned event data.
For event selection, we use an extraction circle of
64\arcsec\ (26 pixels) radius and in the 3--50~keV energy range.
Barycentric correction is done using  \texttt{barycorr}.
We use PRESTO and \texttt{prepfold} to perform a pulsation search
and to determine the spin frequency of \psreightone.

\subsection{Chandra data} \label{sec:chandra}

Chandra observed \psreightone\ using the ACIS-S detector in
continuous clocking (CC) mode on 2021 February 10 (ObsID 23545)
and 2021 June 23 (ObsID 23546) for 20~ks on each date
(see Table~\ref{tab:data}).
We reprocess these data following the standard procedure with
\texttt{chandra\_repro} of the Chandra Interactive Analysis of
Observations (CIAO) package version 4.14 and Calibration Database
(CALDB) 4.9.6 \citep{fruscioneetal06}.
We extract events from the one-dimensional CC data sets along a
3\arcsec\ (3.1 pixels) length centered on the pulsar position
(see Section~\ref{sec:1813}) and in the 2.5--8~keV energy range.
We transform the selected event time stamps from TT to TDB using the
\texttt{axbary} tool and the pulsar position calculated at the epoch
of each observation with the proper motion reported in \cite{dzibrodriguez21}.
As with analysis of NICER data of the other pulsars,
we use PRESTO and \texttt{prepfold} to perform a pulsation search
and to determine the spin frequency of \psreightone.
\vspace{-1em}

\section{Results} \label{sec:results}

\subsection{\psrzero} \label{sec:0058}

\begin{figure}
\plotone{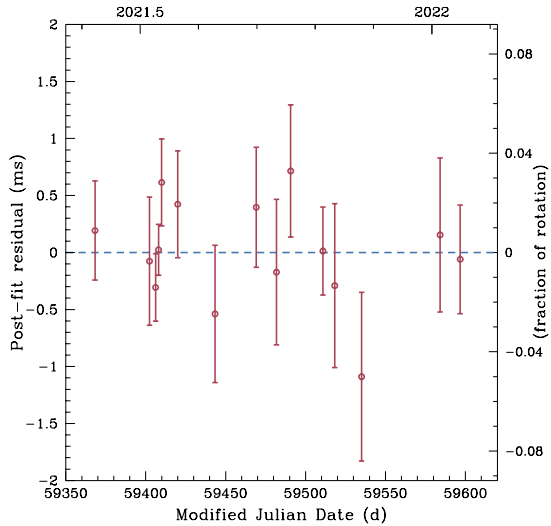}
\caption{Timing residuals of \psrzero\ from a best-fit of
NICER pulse times-of-arrival with the timing model given in
Table~\ref{tab:0058}.
Errors are 1$\sigma$ uncertainty.
\label{fig:0058tres}}
\end{figure}

NICER observations of \psrzero\ began on 2021 June 1, and we are
able to obtain a phase-connected timing model using data through
2022 January 25.
Figure~\ref{fig:0058tres} shows timing residuals of the 15 TOAs used
to obtain our best-fit timing model, which is given in Table~\ref{tab:0058}.
While the initial timing model only has a precision of 10\%
in $\nudot$ \citep{maitraetal21}, the new longer timespan data
yield a precision of 0.002\% in $\nudot$,
as well as a $\nuddot$ and thus braking index $n=50\pm2$.
\vspace{-1em}

\begin{deluxetable}{lc}
\tablecaption{Timing parameters of \psrzero\label{tab:0058}}
\tablehead{
\colhead{Parameter} & \colhead{Value}
}
\startdata
R.A. $\alpha$ (J2000) & $00^{\rm h}58^{\rm m}16^{\rm s}\!\!.85$ \\
Decl. $\delta$ (J2000) & $-72^\circ18\arcmin05\farcs60$ \\
Solar system ephemeris & DE405 \\
Range of dates (MJD) & 59366$-$59604 \\
Epoch $t_0$ (MJD TDB) & 59408 \\
Frequency $\nu$ (Hz) & 45.940434278(6) \\
Freq.\ 1st derivative $\nudot$ (Hz s$^{-1}$) & $-6.2324(1)\times10^{-11}$ \\
Freq.\ 2nd derivative $\nuddot$ (Hz s$^{-2}$) & $4.2(2)\times10^{-21}$ \\
RMS residual ($\mu$s) & 352 \\
$\chi^2$/dof & 9.94/11 \\
Number of TOAs & 15 \\
\enddata
\tablecomments{
Number in parentheses is 1$\sigma$ uncertainty in last digit.
Position is from a Chandra ACIS-S image (MJD 56332), with a 90\% confidence
level uncertainty of $0\farcs6$ \citep{maitraetal15}.}
\vspace{-2em}
\end{deluxetable}

\begin{figure}
\plotone{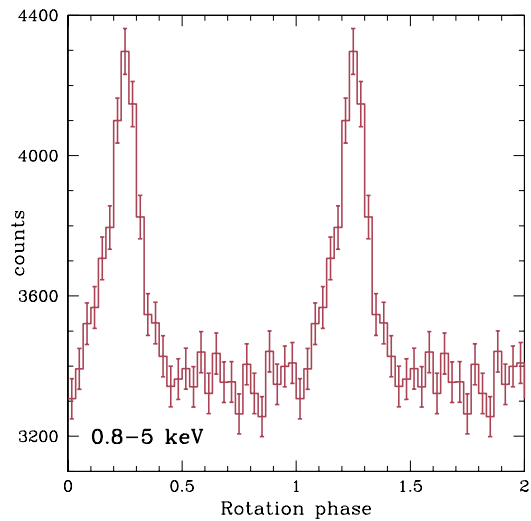}
\caption{Pulse profile of \psrzero\ from 153~ks of NICER data at 0.8--5~keV.
Errors are 1$\sigma$ uncertainty.
Two rotation cycles are shown for clarity.
\label{fig:0058profile}}
\end{figure}

Figure~\ref{fig:0058profile} shows the pulse profile at 0.8--5~keV
from the combined NICER observations.
The pulse profile can be fit by a Gaussian with a full width at
half maximum of 0.14 in phase or 3~ms.
Pulsations are only weakly detected using NICER at 0.3--0.8~keV
and not detected above background at 5--10~keV due to
contamination from the underlying supernova remnant.

No clear spin-up glitches are detected in NICER data so far,
although we can estimate a glitch wait time (albeit applicable
for pulsars with $|\nudot|<3\times10^{-11}\mbox{ Hz s$^{-1}$}$;
\citealt{fuentesetal17})
of $1/(420\mbox{ Hz$^{-1}$}|\nudot|)\approx15\mbox{ months}$
that is longer than our current timespan of observations.
On the other hand, when we extrapolate the spin frequency using the
NICER timing model back to the epoch of the XMM-Newton observation
on 2020 March 15 (MJD 58924), the result is 57.2~$\mu$Hz higher
than that reported in \citet{maitraetal21}, with an uncertainty
of 0.4~$\mu$Hz from the XMM-Newton measurement and 0.2~$\mu$Hz
from our $\nuddot$ uncertainty.
This difference in spin frequency could be due to one or more
spin-up glitches occurring in the 15~months between the times
of the XMM-Newton and NICER observations.
In fact, the large braking index ($n=50$) suggests recovery
from a recent large glitch, similar to behavior seen in \psrfive.
Detection of future large glitches in \psrzero\ would validate
this possibility.

\begin{deluxetable*}{ccccccccccc}
\tablecaption{Timing parameters of \psrfive\label{tab:0537}}
\tablehead{
\colhead{Segment} & \colhead{Epoch} & \colhead{Start} & \colhead{End} & \colhead{$\nu$} & \colhead{$\nudot$} & \colhead{$\nuddot$} & \colhead{$\nig$} & \colhead{RMS} & \colhead{$\chi^2/\mbox{dof}$} & \colhead{TOAs} \\
 & (MJD) & (MJD) & (MJD) & (Hz) & ($10^{-10}\mbox{ Hz s$^{-1}$}$) & ($10^{-20}\mbox{ Hz s$^{-2}$}$) & & ($\mu$s) & &
}
\startdata
11 & 59195 & 59107.7 & 59283.4 & 61.904295359(1) & $-$1.997588(2) & 0.59(2) & 9.2(3) & 179.9 & 8.5 & 31 \\
12 & 59318 & 59286.9 & 59349.9 & 61.902180422(4) & $-$1.99789(2) & 0.7(4) & 11(6) & 82.4 & 2.6 & 10 \\
13 & 59399 & 59352.6 & 59446.6 & 61.900794415(3) & $-$1.997802(9) & 1.6(1) & 25(2) & 132.4 & 4.6 & 17 \\
14 & 59487 & 59461.5 & 59518.4 & 61.899292039(3) & $-$1.99828(3) & [1]\tablenotemark{a} & \nodata & 72.7 & 3.0 & 6 \\
15 & 59578 & 59529.6 & 59626.6 & 61.897742999(2) & $-$1.998105(6) & 1.04(8) & 16(1) & 110.8 & 3.2 & 20 \\
\enddata
\tablecomments{
Columns are interglitch segment number, timing model epoch,
segment start and end dates, spin frequency and its first two time derivatives,
interglitch braking index, timing model residual, goodness-of-fit measure,
and number of times of arrival.
Number in parentheses is 1$\sigma$ uncertainty in last digit.
Segments 1--7 are in \citet{hoetal20}, and
segments 8-10 are in \citet{abbottetal21}.
Position of R.A.=$05^{\rm h}37^{\rm m}47^{\rm s}\!\!.416$,
Decl.=$-69^\circ10\arcmin19\farcs88$ (J2000) is
from a Chandra ACIS-I image (MJD 51442), with 1$\sigma$ uncertainty
of $\sim0\farcs6$ \citep{townsleyetal06}.
Solar system ephemeris used is DE421.
}
\tablenotetext{a}{$\nuddot$ is fixed at $10^{-20}\mbox{ Hz s$^{-2}$}$.}
\vspace{-2em}
\end{deluxetable*}

\begin{deluxetable}{cccccc}
\tablecaption{Glitch parameters of \psrfive\label{tab:glitch}}
\tablehead{
\colhead{Glitch} & \colhead{Glitch epoch} & \colhead{$\Delta\Phi$} & \colhead{$\Delta\nu$} & \colhead{$\Delta\nudot$} & \colhead{$\Delta\nuddot$} \\
 & (MJD) & (cycle) & ($\mu$Hz) & ($10^{-13}\mbox{ Hz s$^{-1}$}$) & ($10^{-20}\mbox{ Hz s$^{-2}$}$)
}
\startdata
11 & 59103(5) & 0.5(6) & 33.9(4) & $-$1(1) & $-$3(2) \\
12 & 59285(2) & $-$0.26(1) & 7.872(8) & $-$0.94(3) & \nodata \\
13 & 59351(2) & 0.51(2) & 12.27(3) & $-$0.8(2) & 0.9(4) \\
14 & 59454(8) & 0.31(1) & 16.60(1) & $-$1.71(4) & \nodata \\
15 & 59522(8) & $-$0.30(2) & 22.08(1) & $-$0.61(3) & \nodata \\
\enddata
\tablecomments{
Columns are glitch number and epoch and
change in rotation phase and changes in spin frequency and its
first two time derivatives at each glitch.
Number in parentheses is 1$\sigma$ uncertainty in last digit.
Glitches 1--7 are in \citet{hoetal20}, and
glitches 8-10 are in \citet{abbottetal21}.
}
\vspace{-4em}
\end{deluxetable}

\subsection{\psrfive} \label{sec:0537}

\psrfive\ has been observed by NICER since early in the start of
the mission in mid-2017.
Data on \psrfive\ from 2017 August 17 to 2020 April 25 and their timing
results, including eight glitches during this period, are reported
in \citet{hoetal20}.
Data from 2020 May 12 to October 29 and their results, including three
more glitches during this period, are reported in \citet{abbottetal21}.
Here we report on data from 2020 November 10 to 2022 February 17.
During this period, four new glitches are detected.
The timing model and glitch parameters are given in
Tables~\ref{tab:0537} and \ref{tab:glitch}, respectively.
We use the previous naming convention where each segment is separated
by a glitch and is labeled by glitch number, with segment 1 occurring
after glitch 1, which is the first NICER-detected glitch.
Note that glitch 8 is first reported in \citet{hoetal20}, but
revised timing parameters for segment 8 and glitch 8 are given
in \citet{abbottetal21} after accumulation of more data for the segment;
similarly here we revise parameters first given in \citet{abbottetal21}
for segment 11 and glitch 11.
We find the epoch of glitch 15 to be MJD 59522, even though a glitch
epoch between MJD 59529 and 59556 produces similar fit results; we
favor the earlier epoch since periodicity analyses of an observation
on MJD 59529 suggest the glitch already occurred by this date.

\begin{figure}
\plotone{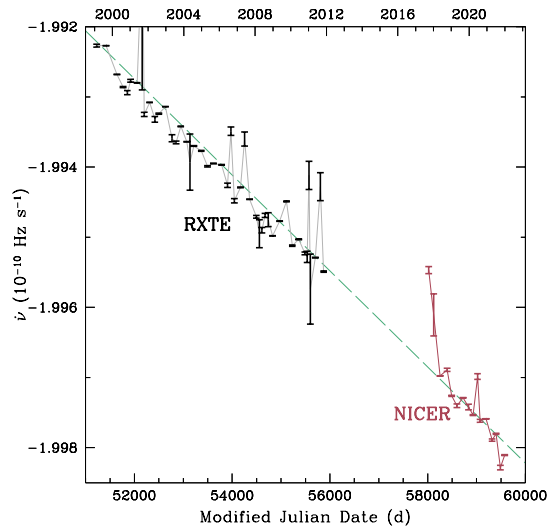}
\caption{Evolution of spin frequency time derivative $\nudot$ of \psrfive.
$\nudot$ are measured by fitting a timing model to TOAs in each
interglitch segment
(see Table~\ref{tab:0537} and \citealt{hoetal20,abbottetal21} for NICER
and Table~1 of \citealt{antonopoulouetal18} for RXTE).
Errors are 1$\sigma$ uncertainty.
Dashed line shows a linear fit of NICER and RXTE data with
best-fit $\nuddot=-7.92\times 10^{-22}\mbox{ Hz s$^{-2}$}$.
\label{fig:f1}}
\end{figure}

Figure~\ref{fig:f1} shows $\nudot$ for each segment, as well as
interglitch $\nudot$ values measured using RXTE from
\citet{antonopoulouetal18}.
A simple linear fit to only the NICER set of $\nudot$ gives
$\nuddot=(-8.2\pm0.5)\times10^{-22}\mbox{ Hz s$^{-2}$}$
and a long-term braking index $n=-1.27\pm0.08$,
while a fit to the entire NICER and RXTE set of $\nudot$ gives
$\nuddot=(-7.92\pm0.06)\times10^{-22}\mbox{ Hz s$^{-2}$}$
and $n=-1.234\pm0.009$;
all are in agreement with values found in \citet{hoetal20} and
values found using only RXTE data in \citet{antonopoulouetal18,ferdmanetal18}.
Thus there is not strong evidence for a change in the long-term
braking index of \psrfive.

\begin{figure}
\plotone{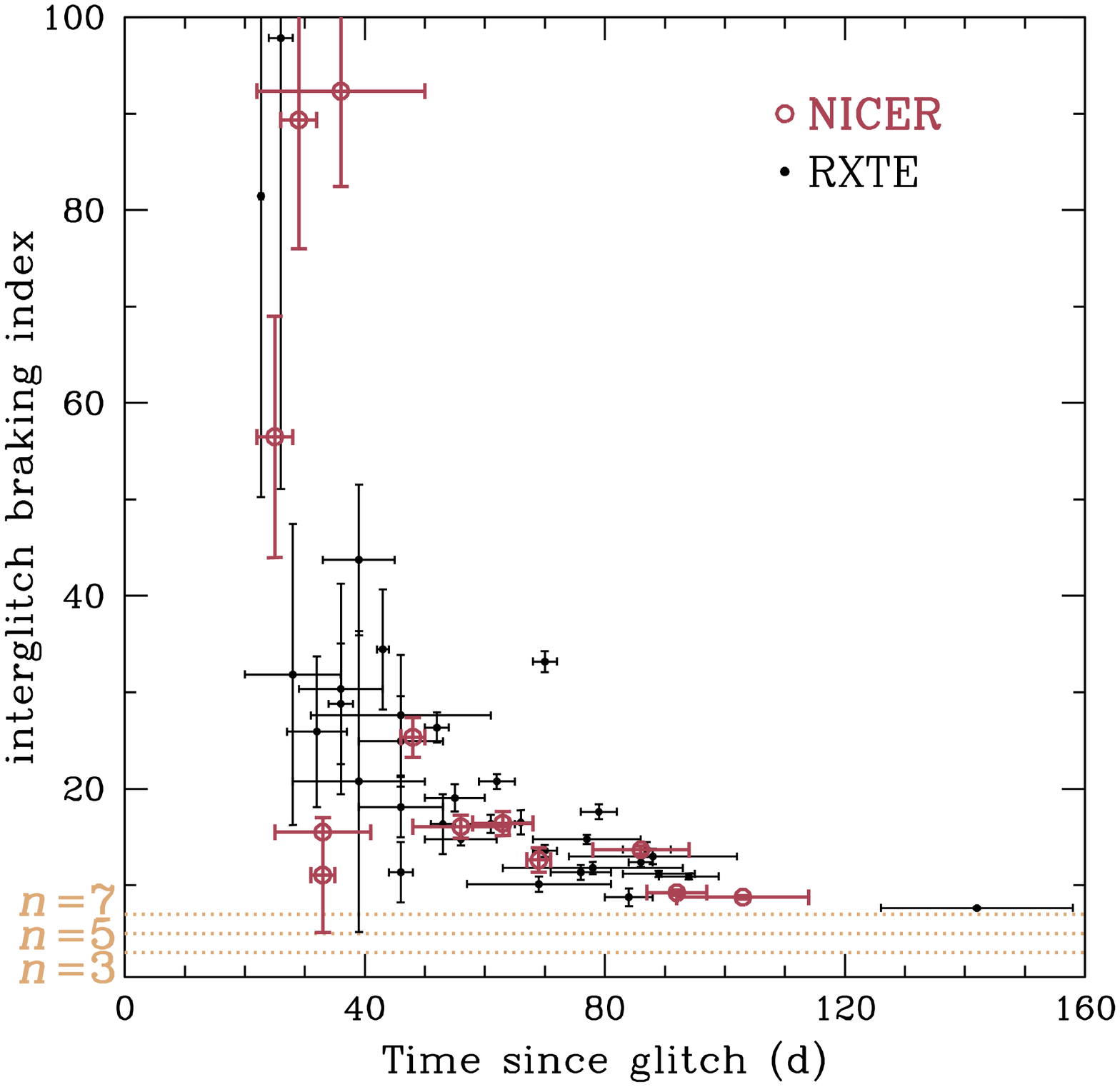}
\caption{Interglitch braking index $n_{\rm ig}$ of \psrfive\ calculated
from spin parameters of each segment between glitches as a
function of time since the last glitch.
Large and small circles denote NICER and RXTE values, respectively
(from here and \citealt{antonopoulouetal18,hoetal20,abbottetal21}).
Errors in $n_{\rm ig}$ are 1$\sigma$.
Horizontal dotted lines indicate braking index $n=3$, 5, and 7,
which are expected for pulsar spin-down by electromagnetic dipole
radiation, gravitational wave-emitting mountain, and
gravitational wave-emitting r-mode oscillation, respectively.
\label{fig:nig}}
\end{figure}

As discussed in \citet{hoetal20} (see also
\citealt{middleditchetal06,anderssonetal18,antonopoulouetal18,ferdmanetal18}),
the short-term spin-down behavior (i.e., behavior in between glitches)
is much different than the long-term behavior.
In particular, the interglitch braking index $\nig$ is non-negative
and much greater than the canonical value of 3 that is the result of
conventional spin-down by electromagnetic dipole radiation at constant
magnetic field and moment of inertia.
This is illustrated in Figure~\ref{fig:nig}, which shows
$\nig$ measured using NICER and RXTE, with the latter values taken
from \citet{antonopoulouetal18},
and time since last glitch is the epoch of the segment minus the epoch
of the corresponding glitch
(e.g., time since glitch 15 $=59195-59103=92\mbox{ d}$).
It is clear that large values of $\nig$ are measured for short
times after a glitch
and that small $\nig$ are measured after long post-glitch times.
In other words, it appears there is recovery back to a rotational
behavior that is characterized by a braking index $\lesssim7$
after disruption by a glitch.
Fits to an exponential decay yield decay timescales of 19--44~d,
with a longer timescale for a lower asymptotic braking index \citep{hoetal20}.
Braking indices of 5 and 7 are expected for spin-down by gravitational
wave quadrupole and r-mode emission, respectively
(see \citealt{hoetal20,abbottetal21}, and references therein).

\begin{figure}
\plotone{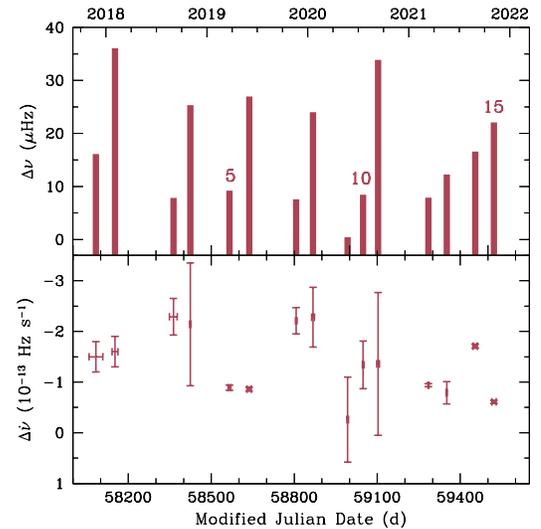}
\caption{
Glitch $\Delta\nu$ (top) and $\Delta\nudot$ (bottom) as functions of time
(see Table~\ref{tab:glitch} and \citealt{hoetal20,abbottetal21}).
Errors in $\Delta\nudot$ are 1$\sigma$ uncertainty.
\label{fig:0537glitch}}
\end{figure}

Figure~\ref{fig:0537glitch} shows glitch parameters $\Delta\nu$
and $\Delta\nudot$ for the 15 glitches measured so far using
NICER (see Table~\ref{tab:glitch} and \citealt{hoetal20,abbottetal21}).
Note that the alternating sizes of $\Delta\nu$ and pairing of $\Delta\nudot$
seen in the first 8 glitches does not continue in more recent glitches.

\begin{figure}
\plotone{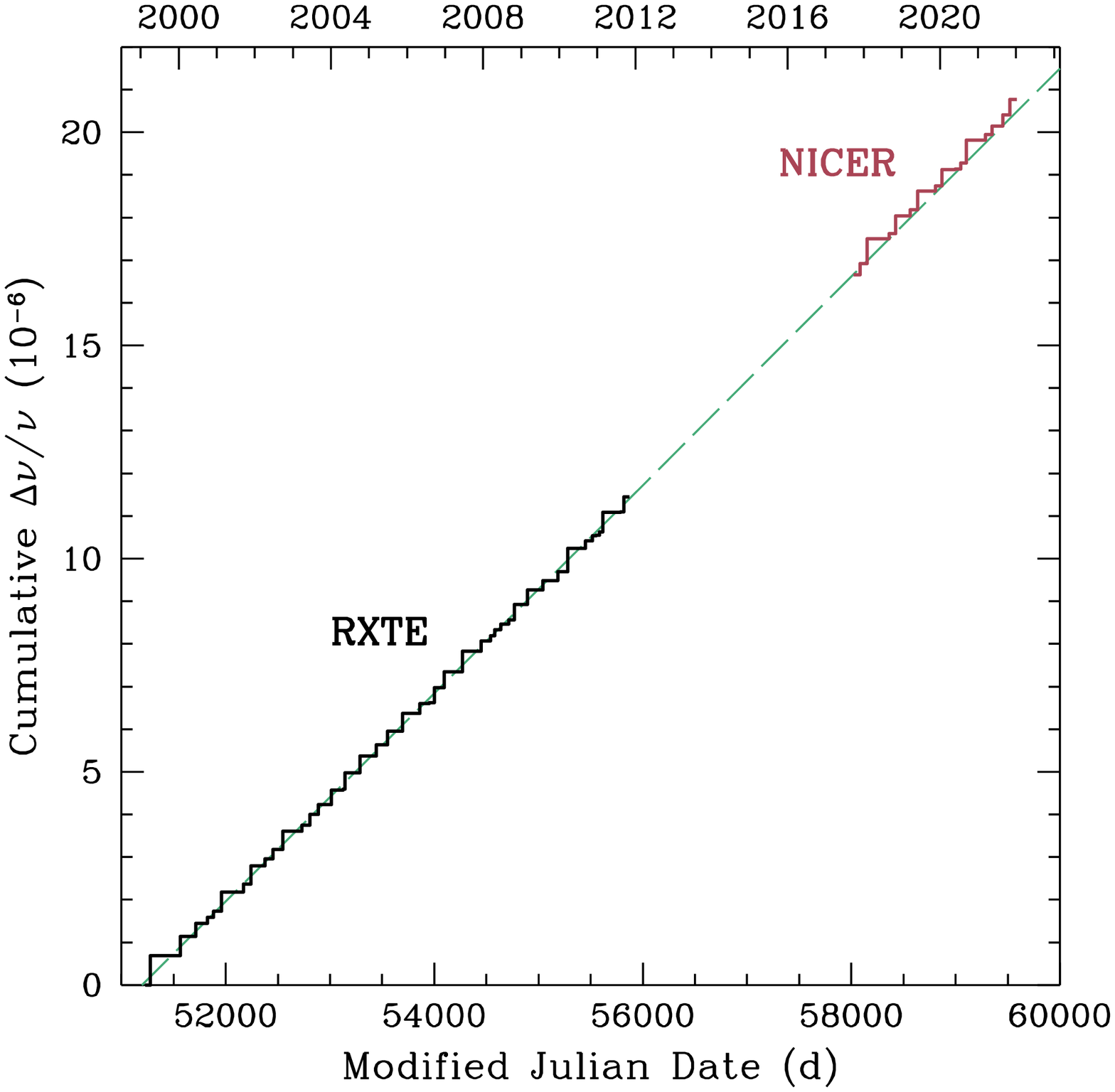}
\caption{
Fractional glitch magnitude $\Delta\nu/\nu$ of \psrfive\ shown as a 
cumulative sum over each previous glitch.
RXTE values are from Table~2 of \citet{antonopoulouetal18}.
Dashed line indicates a line with a slope of
$8.918\times 10^{-7}\mbox{ yr$^{-1}$}$, which is the glitch activity
$A_{\rm g}\equiv\sum_i(\Delta\nu/\nu)_i/\tobs$, where
$\tobs$ is time over which the pulsar is monitored.
NICER values are offset by $\Delta\nu/\nu=16.7\times10^{-6}$,
which is the extrapolated value of RXTE-only glitch activity
at epoch of NICER segment 0 at MJD 58020.
\label{fig:activity}}
\end{figure}

Glitch activity can be characterized by the parameter
$\Ag\equiv\sum_i(\Delta\nu/\nu)_i/\tobs$, where the summation is over
each glitch $i$ and $\tobs$ is time over which the pulsar is
monitored \citep{mckennalyne90}.
For glitches detected using NICER and $\tobs=4.5\mbox{ yr}$,
we find $\sum_i\Delta\nu_i=(254.6\pm0.6)\mbox{ $\mu$Hz}$ and
$\Ag=(9.15\pm0.02)\times 10^{-7}\mbox{ yr$^{-1}$}$.
Figure~\ref{fig:activity} plots the cumulative fractional glitch
magnitude $\Delta\nu/\nu$ over the RXTE and NICER eras.
Combining RXTE and NICER glitches produces an activity parameter
$\Ag=(8.918\pm0.009)\times 10^{-7}\mbox{ yr$^{-1}$}$.

\begin{figure}
\plotone{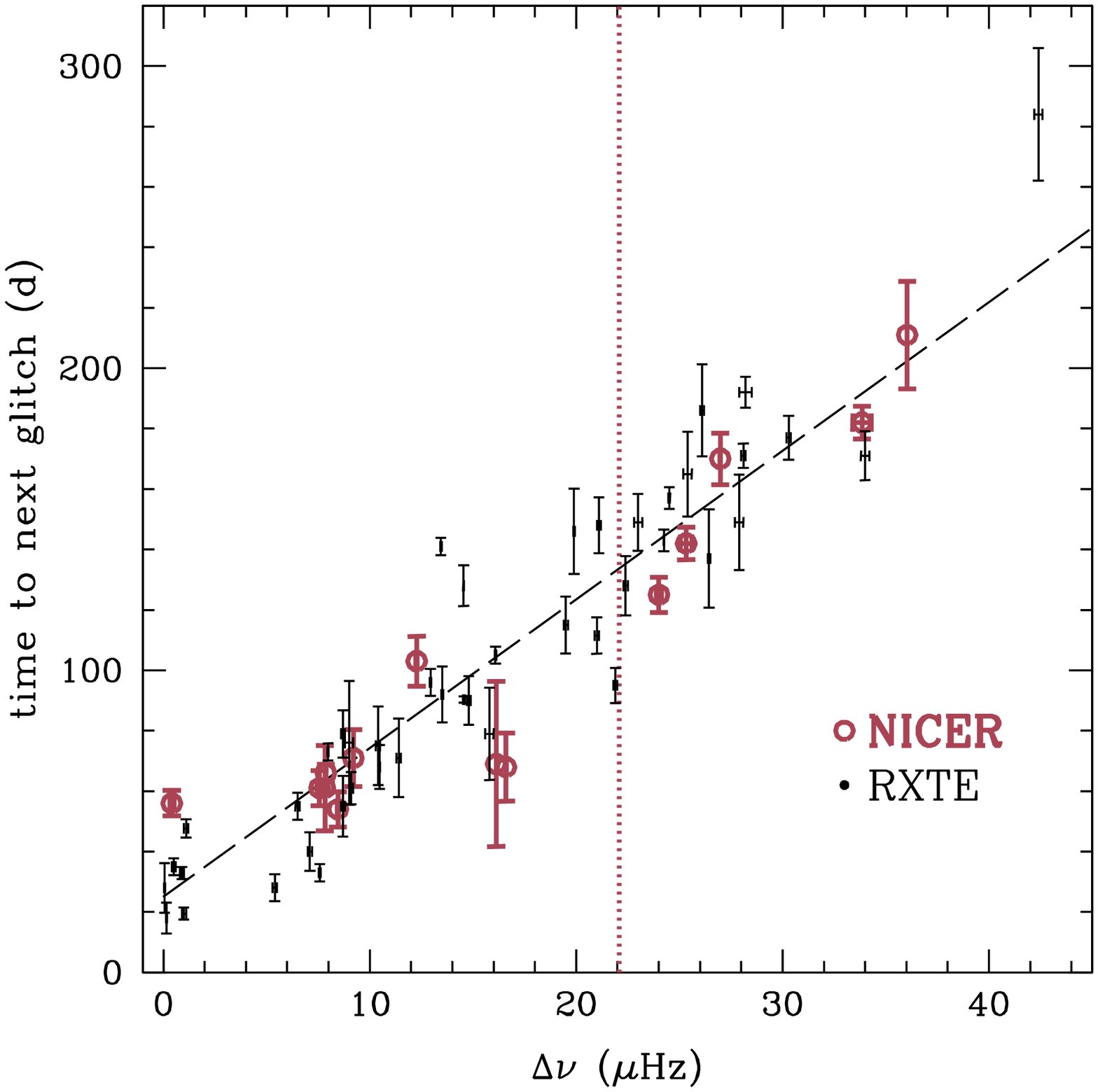}
\caption{Correlation between time interval to
the next glitch $\Delta T$ and size of glitch $\Delta\nu$ of \psrfive.
Large and small circles denote NICER and RXTE values, respectively
(from here and \citealt{antonopoulouetal18,hoetal20,abbottetal21}).
Errors in $\Delta\nu$ are 1$\sigma$.
The vertical dotted line indicates the size of NICER glitch 15,
which is the most recent glitch (on 2021 November 4) and for which
time to next glitch is not known yet.
Dashed line shows linear fit result
$\Delta T=49.2\mbox{ d }(\Delta\nu/\mbox{10 $\mu$Hz})+25\mbox{ d}$.
\label{fig:glitchpredict}}
\end{figure}

Glitches of \psrfive\ are unique in that the time to next glitch is
correlated with the size of the preceding glitch (see also
\citealt{middleditchetal06,antonopoulouetal18,ferdmanetal18,hoetal20}).
This is illustrated in Figure~\ref{fig:glitchpredict}.
The correlation can be fit by time to
\mbox{next glitch $=49.2\mbox{ d }(\Delta\nu/\mbox{10 $\mu$Hz})+(25\pm18)\mbox{ d}$},
in agreement with that found in \citet{hoetal20}.
This correlation enables prediction of when glitches will occur in
\psrfive.
In particular, glitch 16 should occur on 2022 March 18, although there
is a large uncertainty ($\pm 26\mbox{ d}$) due in part to the
uncertain epoch of glitch 15.

\subsection{\psrone} \label{sec:1101}

\begin{figure}
\plotone{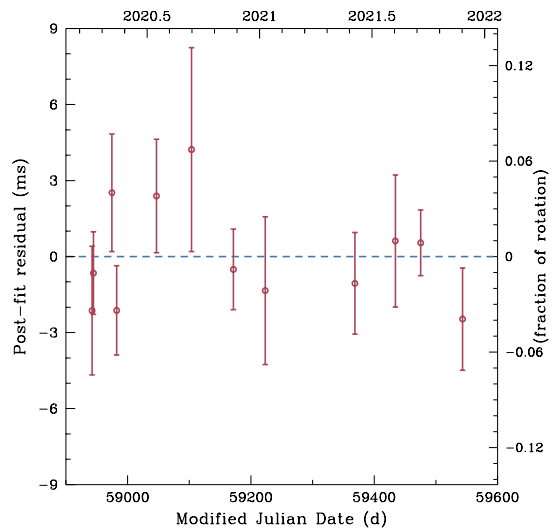}
\caption{Timing residuals of \psrone\ from a best-fit of
NICER pulse times-of-arrival with the timing model given in
Table~\ref{tab:1101}.
Errors are 1$\sigma$ uncertainty.
\label{fig:1101tres}}
\end{figure}

\begin{figure}
\plotone{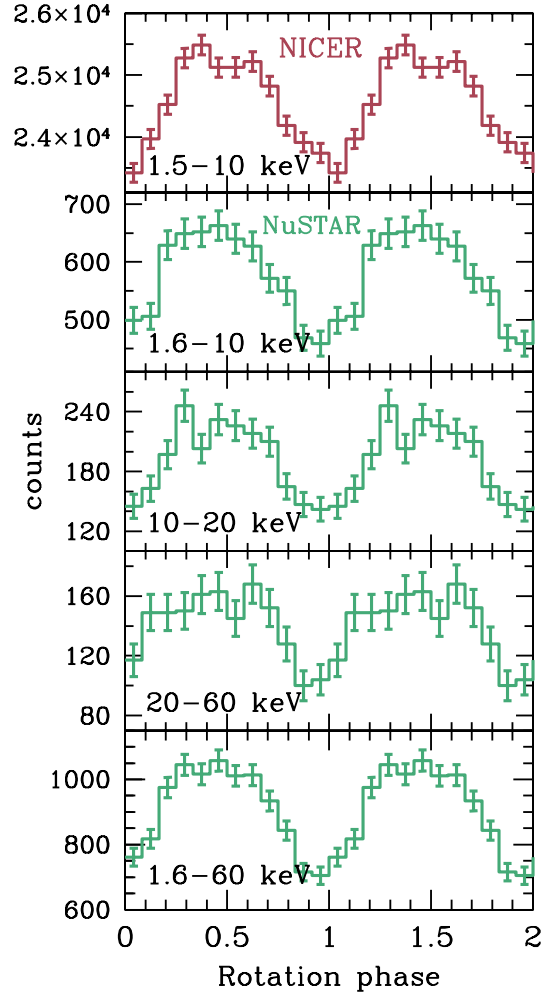}
\caption{Pulse profile of \psrone\
from 370~ks of NICER data at 1.5--10~keV (top),
while lower panels show pulse profiles from 136~ks of NuSTAR data at
1.6--10~keV, 10--20~keV, 20--60~keV, and 1.6--60~keV.
Errors are 1$\sigma$ uncertainty.
Two rotation cycles are shown for clarity.
\label{fig:1101profile}}
\end{figure}

NICER observations of \psrone\ began on 2020 April 1, and we are
able to obtain a phase-connected timing model using data through
2021 December 16.
Figure~\ref{fig:1101tres} shows timing residuals of the 12 TOAs used
to obtain our best-fit timing model, which is given in Table~\ref{tab:1101},
and Figure~\ref{fig:1101profile} shows the 1.5--10~keV pulse profile
from the combined NICER observations.
Our measured spin-down rate of
$\nudot=(-2.26504\pm0.0004)\times10^{-12}\mbox{ Hz s$^{-1}$}$
is consistent with and significantly improves upon the precision
of the previously measured incoherent timing model value of
$\nudot=(-2.17\pm0.13)\times10^{-12}\mbox{ Hz s$^{-1}$}$
from \citet{halpernetal14}.
The addition of $\nuddot$ to the timing model yields a fit
improvement of only $\Delta\chi^2=0.5$ and an unconstrained
$\nuddot=(-1\pm18)\times10^{-24}\mbox{ Hz s$^{-2}$}$.

\begin{deluxetable}{lc}
\tablecaption{Timing parameters of \psrone\label{tab:1101}}
\tablehead{
\colhead{Parameter} & \colhead{Value}
}
\startdata
R.A. $\alpha$ (J2000) & $11^{\rm h}01^{\rm m}44^{\rm s}\!\!.915$ \\
Decl. $\delta$ (J2000) & $-61^\circ01\arcmin38\farcs66$ \\
Solar system ephemeris & DE405 \\
Range of dates (MJD) & 58940$-$59564 \\
Epoch $t_0$ (MJD TDB) & 59171 \\
Frequency $\nu$ (Hz) & 15.9230402557(5) \\
Freq.\ 1st derivative $\nudot$ (Hz s$^{-1}$) & $-2.26504(4)\times10^{-12}$ \\
RMS residual (ms) & 1.60 \\
$\chi^2$/dof & 7.80/9 \\
Number of TOAs & 12 \\
\enddata
\tablecomments{
Number in parentheses is 1$\sigma$ uncertainty in last digit.
Position is from a Chandra ACIS-I image (MJD 56211), with a 90\% confidence
level uncertainty of $0\farcs64$ \citep{pavanetal14}.
}
\end{deluxetable}

The timespan of our NICER timing model overlaps with a NuSTAR
observation taken on 2020 November 20.
We use the NICER timing model to extract pulsed emission from
the NuSTAR data.  The resulting pulse profiles are shown in
Figure~\ref{fig:1101profile}.  It is clear that the two sets of pulse
profiles bear strong resemblance in shape and have consistent alignment.
More detailed analysis of the NuSTAR data can be found in
\citet{klingleretal22}.

\subsection{\psrfour} \label{sec:1412}

\begin{figure}
\plotone{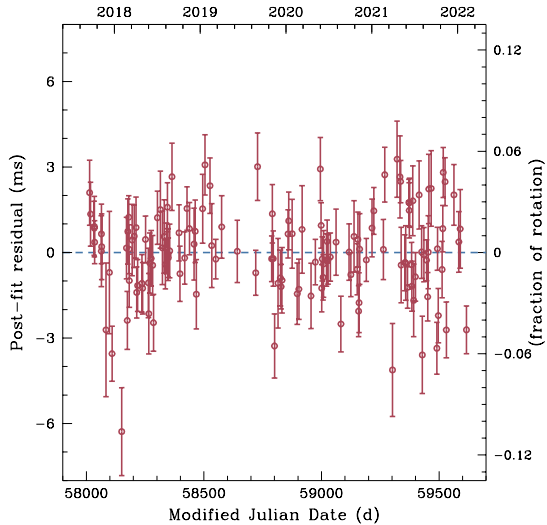}
\caption{Timing residuals of \psrfour\ from a best-fit of
NICER pulse times-of-arrival with the timing model given in
Table~\ref{tab:1412}.
Errors are 1$\sigma$ uncertainty.
\label{fig:1412}}
\end{figure}

NICER observations of \psrfour\ began on 2017 September 15, and
\citet{bogdanovetal19} report a 1~yr phase-connected timing model
using NICER data through 2018 October 3.
\citet{mereghettietal21} report a 3.4~yr phase-connected timing model
using NICER data through 2021 February 26.
Here we extend the timing model to over 4.4~yr using NICER data
through 2022 February 8.
Our analysis procedure yields 138 TOAs that are barycentered, but
not corrected for proper motion, with respect to the pulsar's
position as measured by \citet{halperngotthelf15}.
While we are able to successfully obtain a phase-connected timing model,
the resulting timing residuals over the 4.4~yr span of data
display a systematic wave-like behavior on a several months timescale,
even with the addition of $\nuddot$ and $\dddot{\nu}$ terms in the timing model
(see, e.g., Figure~1 of \citealt{mereghettietal21}).
Therefore, we consider a timing model that includes the pulsar's
position as a fit parameter.  To construct such a model, we calculate
138 spacecraft topocentric, not barycentric, TOAs from the same dataset,
and we fit these TOAs using PINT \citep{luoetal21}.
Figure~\ref{fig:1412} shows timing residuals of the TOAs used
to obtain our best-fit timing model, which is given in Table~\ref{tab:1412}.
While the timing model of \citet{bogdanovetal19} only required a
spin-down $\nudot$ term to achieve a timing residual of 1.36~ms,
our four times longer time baseline requires $\nuddot$ for a
comparable timing residual of 1.45~ms.
The addition of proper motion as a fit parameter to the timing model
only yields a fit improvement of $\Delta\chi^2=13$.

\begin{deluxetable}{lc}
\tablecaption{Timing parameters of \psrfour\label{tab:1412}}
\tablehead{
\colhead{Parameter} & \colhead{Value}
}
\startdata
R.A. $\alpha$ (J2000) & $14^{\rm h}12^{\rm m}56^{\rm s}\!\!.05(3)$ \\
Decl. $\delta$ (J2000) & $+79^\circ22\arcmin03\farcs68(7)$ \\
Solar system ephemeris & DE405 \\
Range of dates (MJD) & 58014.2$-$59616.9 \\
Epoch $t_0$ (MJD TDB) & 58750 \\
Frequency $\nu$ (Hz) & 16.8921082712(1) \\
Freq.\ 1st derivative $\nudot$ (Hz s$^{-1}$) & $-9.40547(4)\times10^{-13}$ \\
Freq.\ 2nd derivative $\nuddot$ (Hz s$^{-2}$) & $-2.83(3)\times10^{-23}$ \\
RMS residual (ms) & 1.449 \\
$\chi^2$/dof & 303.7/132 \\
Number of TOAs & 138 \\
\enddata
\tablecomments{Number in parentheses is 1$\sigma$ uncertainty in last digit.
Position epoch is the same as timing model epoch.
No proper motion is assumed.}
\vspace{-1em}
\end{deluxetable}

It is notable that the position we measure using timing data is
different from that measured by \citet{halperngotthelf15} using
Chandra HRC-I imaging data from 2007 and 2014.
In particular, their position from the 2014 data (MJD 56749),
which is much longer and closer to the time of our data than the 2007 data,
is ($\alpha$,$\delta$)=
($14^{\rm h}12^{\rm m}55^{\rm s}\!\!.815\pm0^{\rm s}\!\!.011$,
$+79^\circ22\arcmin03\farcs697\pm0\farcs030$),
and thus our position differs by
$\Delta\alpha=+0^{\rm s}\!\!.23\pm0^{\rm s}\!\!.04$
and $\Delta\delta=-0\farcs02\pm0\farcs10$.
The implied proper motion of \psrfour\ between its 2014 and 2019
positions is $\mu_\alpha\cos\delta=+120\pm20\mbox{ mas yr$^{-1}$}$
and $\mu_\delta=-3\pm20\mbox{ mas yr$^{-1}$}$, which is also at odds
with $\mu_\alpha\cos\delta=-40\pm30\mbox{ mas yr$^{-1}$}$ and
$\mu_\delta=-56\pm21\mbox{ mas yr$^{-1}$}$ as measured by
\citet{halperngotthelf15} from changes of the 2007 and 2014 positions.
Differences in the two techniques used, i.e., imaging versus timing,
could contribute to the different measured positions of \psrfour,
including the impact of timing noise on the timing model fits.
A future Chandra observation may resolve these discrepancies.

\subsection{\psreightone} \label{sec:1813}

Our analysis of \psreightone\ only considers the spin frequencies
measured using various radio and X-ray telescopes over a 12~yr timespan.
In particular, we fit the evolution of $\nu$ with a linear decline
in order to determine $\nudot$.
This is done because data for \psreightone\ are sparse and a
phase-connected timing analysis like that done in \citet{hoetal20a}
requires significant NICER observing time.

\begin{figure}
\plotone{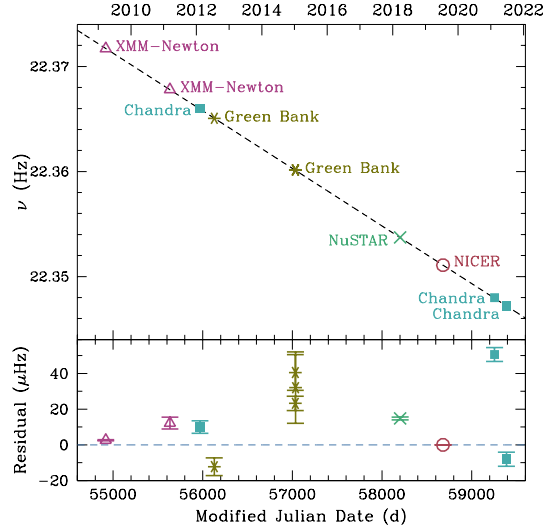}
\caption{Spin frequency $\nu$ of \psreightone\ (top) and difference between
best-fit linear model and data (bottom) as functions of time.
Measurements of $\nu$ are made using XMM-Newton (triangles), Chandra (squares),
Green Bank Telescope in radio (stars), NuSTAR (cross), and NICER (circle).
Dashed line shows a linear fit of all $\nu$ measurements with
best-fit $\nudot=-6.3442\times10^{-11}\mbox{ Hz s$^{-1}$}$.
\label{fig:1813}}
\end{figure}

We obtain three new measurements of the spin frequency of \psreightone.
In particular,
we find $\nu=22.3537223\pm0.0000008\mbox{ Hz}$ on MJD~58202.39
from the 2018 NuSTAR observation
and $\nu=22.3479859\pm0.0000039\mbox{ Hz}$ on MJD~59255.45
and $\nu=22.3471999\pm0.0000039\mbox{ Hz}$ on MJD~59388.14
from the two 2021 Chandra continuous clocking observations.
Meanwhile, previous spin frequency measurements are from
2009 and 2011 using XMM-Newton and 2012 using Chandra \citep{halpernetal12},
2012 and 2015 measurements at radio wavelengths \citep{camiloetal21}, and
2019 measurement using NICER \citep{hoetal20a}.
These are all shown in Figure~\ref{fig:1813}.
Fitting a simple linear decline in spin frequency yields a best-fit
spin-down rate $\nudot=(-6.34425\pm0.00072)\times10^{-11}\mbox{ Hz s$^{-1}$}$,
which is in agreement with that determined in \citet{hoetal20a}
because the fit is driven by the two most precise measurements
by XMM-Newton in 2009 and NICER in 2019.

Table~\ref{tab:1813} presents the resulting 12~yr timing model,
albeit one that is incoherent in contrast to the 37~d
phase-connected timing model presented in \citet{hoetal20a}.
Residuals from the timing model suggest \psreightone\ undergoes
glitches with sizes as large as a few tens of $\mu$Hz
(see bottom panel of Figure~\ref{fig:1813}), and the spin-down
rate suggests a glitch wait time of $\sim 14$~months
(\citealt{fuentesetal17}; see Section~\ref{sec:0058}).
Such large glitches are uncommon but can occur multiple times in
a single pulsar over a period of a few years, such as in the
Vela pulsar and \psrfive\ (see, e.g., Figure~\ref{fig:0537glitch}).
Analysis of the 2019 NICER data of \psreightone\ indicates a
spin-down rate of $-6.428\times10^{-11}\mbox{ Hz s$^{-1}$}$ during
the 37~d observing period and a
possible glitch with $\Delta\nu\approx3\mbox{ $\mu$Hz}$ near the end
of the observation (see \citealt{hoetal20a}, for more discussion).
Note that the spin frequency difference of $786\mbox{ $\mu$Hz}$
between the two 2021 Chandra observations separated by 133~d implies
a spin-down rate of $-6.84\times10^{-11}\mbox{ Hz s$^{-1}$}$.

\begin{deluxetable}{lc}
\tablecaption{Incoherent timing parameters of \psreightone\label{tab:1813}}
\tablehead{
\colhead{Parameter} & \colhead{Value}
}
\startdata
R.A., $\alpha$ (J2000) & $18^{\rm h}13^{\rm m}35^{\rm s}\!\!.173$ \\
Decl., $\delta$ (J2000) & $-17^\circ49\arcmin57\farcs75$ \\
Solar system ephemeris & DE405 \\
Range of dates (MJD) & 54918.14$-$59388.14 \\
Epoch $t_0$ (MJD TDB) & 58681.04 \\
Frequency $\nu$ (Hz) & 22.35108384(2) \\
Freq.\ 1st derivative $\nudot$ (Hz s$^{-1}$) & $-6.3442(7)\times10^{-11}$ \\
Proper motion $\mu_\alpha\cos\delta$ (mas yr$^{-1}$) & $-5.0$ \\
Proper motion $\mu_\delta$ (mas yr$^{-1}$) & $-13.2$ \\
\enddata
\tablecomments{
Number in parentheses is 1$\sigma$ uncertainty in last digit.
Position is from VLA data (MJD 58119), with uncertainties of
$\sim0^{\rm s}\!\!.009$ and $\sim0\farcs13$, and
proper motion has uncertainties of 3.7 and $6.7\mbox{ mas yr$^{-1}$}$
in $\mu_\alpha\cos\delta$ and $\mu_\delta$, respectively
\citep{dzibrodriguez21}.
}
\end{deluxetable}

\subsection{\psreight} \label{sec:1849}

\begin{figure}
\plotone{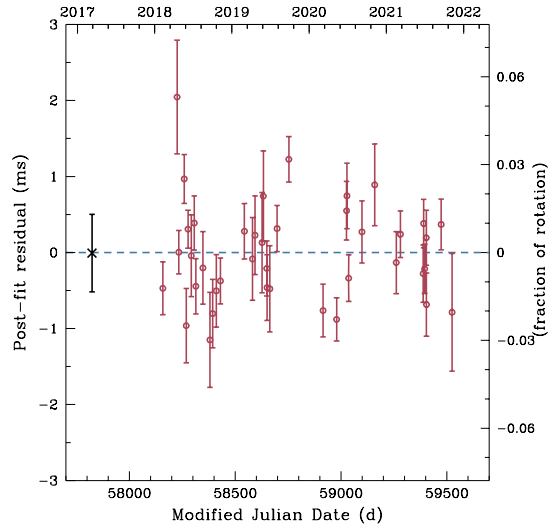}
\caption{Timing residuals of \psreight\ from a best-fit of
NICER (circles) and one Swift (cross) pulse times-of-arrival,
with the timing model given in Table~\ref{tab:1849}.
Errors are 1$\sigma$ uncertainty.
\label{fig:1849}}
\end{figure}

\begin{deluxetable}{lc}
\tablecaption{Timing parameters of \psreight\label{tab:1849}}
\tablehead{
\colhead{Parameter} & \colhead{Value}
}
\startdata
R.A., $\alpha$ (J2000) & $18^{\rm h}49^{\rm m}01^{\rm s}\!\!.632$ \\
Decl., $\delta$ (J2000) & $-00^\circ01\arcmin17\farcs45$ \\
Solar system ephemeris & DE421 \\
Range of dates (MJD) & 57832.1$-$59533.9 \\
Epoch $t_0$ (MJD TDB) & 58682 \\
Frequency $\nu$ (Hz) & 25.9586535424(3) \\
Freq.\ 1st derivative $\nudot$ (Hz s$^{-1}$) & $-9.53597(1)\times10^{-12}$ \\
Freq.\ 2nd derivative $\nuddot$ (Hz s$^{-2}$) & $8.3(1)\times10^{-23}$ \\
Freq.\ 3rd derivative $\dddot{\nu}$ (Hz s$^{-3}$) & $-2.25(3)\times10^{-30}$ \\
Freq.\ 4th derivative $\ddddot{\nu}$ (Hz s$^{-4}$) & $6.6(5)\times10^{-38}$ \\
RMS residual ($\mu$s) & 569 \\
$\chi^2$/dof & 90.7/35 \\
Number of TOAs & 41 \\
\enddata
\tablecomments{
Number in parentheses is 1$\sigma$ uncertainty in last digit.
Position is from a Chandra HRC-S image (MJD 55885), with a 90\% confidence
level uncertainty of $0\farcs6$ \citep{kuiperhermsen15}.
}
\end{deluxetable}

NICER observations of \psreight\ began on 2018 February 13, and
\citet{bogdanovetal19} report a 1.5~yr phase-connected timing model
using NICER data through 2018 September 29 as well as a Swift
observation on 2017 March 19.
Here we extend the timing model to nearly 4.7~yr using NICER data
through 2021 November 16.
Figure~\ref{fig:1849} shows timing residuals of the 41 TOAs used
to obtain our best-fit timing model, which is given in Table~\ref{tab:1849}.
While the timing model of \citet{bogdanovetal19} only required a
spin-down $\nudot$ term to achieve a timing residual of 602~$\mu$s,
our three times longer time baseline requires up to $\ddddot{\nu}$
with a comparable timing residual of 569~$\mu$s.

With our several year time baseline, we attempt to
measure a change of position of \psreight, such as could be caused
by the pulsar's proper motion, using spacecraft topocentric TOAs.
In this case, the best-fit timing model
produces a RMS residual of 513~$\mu$s and
F-test probability of 0.03 and position differences of
$\Delta\alpha=-0^{\rm s}\!\!.008\pm0^{\rm s}\!\!.003$ and
$\Delta\delta=+0\farcs06\pm0\farcs11$
compared to the timing model given in Table~\ref{tab:1849}.
Since the fit improvement is marginal and resulting position and
other timing parameters, e.g., $\nu$ and $\nudot$, are all within
1$\sigma$ uncertainties of the values shown in Table~\ref{tab:1849},
we do not report here this alternative timing model.

\section{Discussion} \label{sec:discuss}

Using long-term monitoring observations made by NICER over the past
several years, we calculated rotation phase-connected timing models
for five pulsars that are only known to be visible at X-ray energies,
as well as determining the long-term spin-down rate of the highly
energetic \psreightone\ by making use of recent Chandra and NuSTAR
observations.
These timing models have a timespan that greatly exceeds that of
previous models, thus providing more reliable characterization of
the spin properties of these rapidly rotating, mostly young pulsars.
Continued monitoring of these pulsars is needed for searches at
other energies (e.g., gamma-ray energies using Fermi; see below)
and crucially for gravitational wave searches of
more sensitive data obtained in upcoming observing runs
(see Section~\ref{sec:intro}).
High-cadence monitoring of \psrfive\ also enables detection of its
glitches, which provides a unique probe of superfluidity.

\subsection{Gamma-ray searches}

The timing models presented here enable us to search for pulsed
gamma-ray emission from \psrzero, \psrone, \psrfour, and \psreight\
in Fermi LAT data.
We neglect \psrfive\ and \psreightone\ because the complex timing
behavior of these pulsars and strong gamma-ray emission from nearby
pulsar wind nebula and supernova remnant make detection of a gamma-ray
pulsar component very difficult (see, e.g., \citealt{ackermannetal15}).
We conduct two types of searches here.
The first follows the methodology described in \citet{smithetal19},
and the second follows that described in \citet{kuiperetal18}.

For the first set of searches,
we use our timing models to gamma-ray phase-fold each pulsar six times,
i.e., using three values of the gamma-ray photon weighting parameter $\mu_w$
described by \citet{bruel19} and used in \citet{smithetal19} and
either LAT data during only the epoch range of each timing model
or all 12.6~yr of currently available LAT data.
We find no statistically significant ($>1\sigma$) deviation from a
uniform phase distribution, and
we place limits on phase-integrated flux above 100~MeV of
$7.2\times10^{-12}\mbox{ erg s$^{-1}$ cm$^{-2}$}$ for \psrone,
$1.8\times10^{-12}\mbox{ erg s$^{-1}$ cm$^{-2}$}$ for \psrfour, and
$1.2\times10^{-11}\mbox{ erg s$^{-1}$ cm$^{-2}$}$ for \psreight.
These limits are based on pulsar sensitivity
estimates made using the second Fermi pulsar catalog (2PC; see in
particular Section~8.2 of \citealt{abdoetal13}) but updated for the
4FGL-DR3 12-year dataset (T.~Burnett, private communication).
For \psrzero, we measure a total flux above 100~MeV of
$(1.6\pm0.3)\times10^{-12}\mbox{ erg s$^{-1}$ cm$^{-2}$}$ that is likely
dominated by emission from other nearby sources
such as the star-forming region NGC~346 (see also \citealt{maitraetal21}).
For \psrone\ and \psreight, detections of their pulsations could be
hindered by several LAT sources that lie within the $1^\circ$ LAT
angular resolution above 500~MeV of each pulsar.
The non-detection of \psrfour, despite its high $\Edot$ relative to
the LAT flux sensitivity at the pulsar's high Galactic latitude and
low gamma-ray background, could be due to an unfavorable viewing
geometry, as illustrated by \citet{johnstonetal20}
and the angles inferred from \citet{mereghettietal21}.
Importantly, the methodology of \citet{smithetal19} may not be
suitable for discovering pulsed emission from soft gamma-ray pulsars.

For the second set of searches using the methodology of \citet{kuiperetal18},
we barycenter LAT timing data of \psrzero\ collected during the
8~month period of the model.
We apply event selections similar to those for the LAT timing
analysis of PSR~J1846$-$0258 presented in \citet{kuiperetal18}.
We then fold in phase the data using the timing parameters
specified in Table~\ref{tab:0058}.
For energies above $\sim 500$~MeV, we obtain a pulse phase
distribution with a narrow single-peaked shape that is coincident in phase
with the prominent pulse in the X-ray band.
However, the overall $Z_n^2$-based significance varies between
2.2--2.8$\sigma$ for $n$ between 3 and 8. Below $\sim 500$~MeV,
the distribution is statistically flat.
More LAT exposure and NICER monitoring are needed to determine
whether this possible detection of pulsed high-energy gamma-rays
from \psrzero\ is real or a statistical fluctuation.

\begin{figure}
\plotone{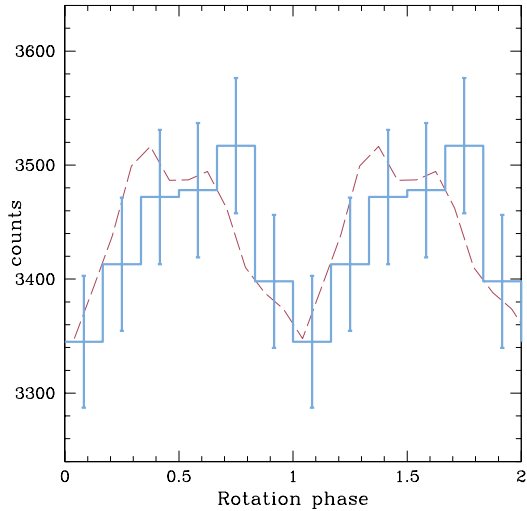}
\caption{Pulse profiles of \psrone\ from Fermi LAT data at 0.3--3~GeV
(solid histogram) and from NICER data at 1.5--10~keV
(dashed curve with arbitrary normalization;
see also Figure~\ref{fig:1101profile}).
Errors are 1$\sigma$ uncertainty.
Two rotation cycles are shown for clarity.
}
\label{fig:1101fermi}
\end{figure}

For \psrone, we use a similar strategy to search LAT data collected
during the 1.7~year period of the ephemeris given in Table~\ref{tab:1101}.
We obtain pulse phase distributions for logarithmically binned
energy bands of 30--300~MeV, 0.3--3~GeV, and 3--30~GeV.
The 30--300~MeV and 3--30~GeV bands
do not show any evidence for a significant pulsed signal.
In the intermediate band, we find a potential signal
at a significance of about $3.2\sigma$ by adopting a $Z_3^2$-test with
the bulk of the enhanced emission (see Figure~\ref{fig:1101fermi})
aligned with the X-ray pulsations detected by NICER and NuSTAR
(see Figure~\ref{fig:1101profile}).
Furthermore, a restricted frequency search in the 0.3--3~GeV band,
near the NICER ephemeris prediction and keeping the frequency
derivative fixed at the value given in Table~\ref{tab:1101}, shows
only one prominent, but weak, maximum coinciding with the prediction
of the NICER ephemeris.
More Fermi LAT exposure and X-ray timing in the future are required
to corroborate this tentative detection of pulsed gamma-ray emission
from \psrone.

For \psrfour\ and \psreight,
we again do not detect pulsed emission using Fermi LAT data.
On the other hand, we detect pulsations of \psreight\ up to
$\sim 150$~keV using the Fermi GBM NaI detectors,
with a $>4\sigma$ significance for the 100--150~keV band
(Kuiper et~al., in prep.).
Note that, even though we detect none (or potentially two) of the
six sources in pulsed gamma-rays, this does not necessarily mean
that each is gamma-ray quiet, given the difficulties described
which hamper possible detection of their pulsations.

\subsection{Superfluidity and glitch predictability from \psrfive} \label{sec:sfglitch}

In the conventional two-component model for glitches
\citep{andersonitoh75,alparetal84},
glitch activity is related to the ratio of superfluid moment of inertia
$I_{\rm sf}$ to total stellar moment of inertia $I$,
such that $I_{\rm sf}/I\ge2\tauc\Ag$
\citep{linketal99}, where $A_{\rm g}\equiv\sum_i(\Delta\nu/\nu)_i/\tobs$;
for simplicity, we neglect here entrainment,
which can be taken into account by a factor of order unity on the
right-hand-side \citep{anderssonetal12,chamel13}.
An equivalent measure of glitch activity to $\Ag$ is
$\nudotg\equiv\sum_i\Delta\nu/\tobs\approx\nu\Ag$ \citep{lyneetal00}
since each glitch induces only a small change in $\nu$,
such that $I_{\rm sf}/I\ge\nudotg/|\nudot|$.
Many works find $2\tauc\Ag$ and $\nudotg/|\nudot|$ can be as large
as $\sim 0.01$ and use this to constrain superfluid properties
and even infer the mass of isolated pulsars
\citep{linketal99,anderssonetal12,chamel13,hoetal15,fuentesetal17,fuentesetal19,montolietal20}.
Note that the mass inferred from glitch activity depends on the
nuclear equation of state, but uncertainty in mass is only weakly
dependent; the mass uncertainty in the case of \psrfive\ is dominated
by uncertainty in its age and hence temperature of its
superfluid/non-superfluid regions (see \citealt{hoetal15}, for details).

\begin{figure}
\plotone{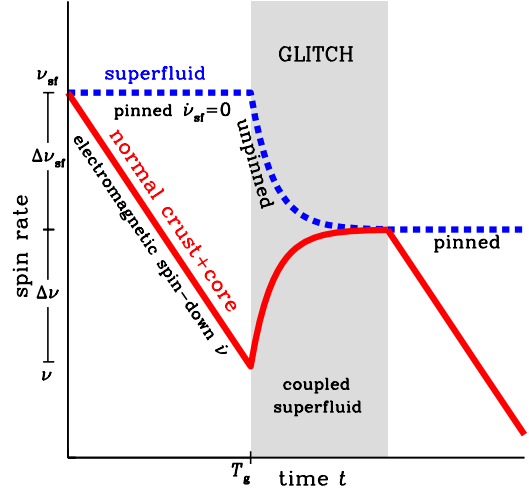}
\caption{Schematic of a glitch (see also \citealt{rayetal19}).
Pulsar spin frequency $\nu$ (solid line)
is observed to decrease at a rate $\nudot$ due to energy loss from
electromagnetic radiation, while a superfluid component within the
star rotating at $\nu_{\rm sf}$ is pinned and does not spin down
with the rest of the star ($\nudot_{\rm sf}=0$; dashed line).
When the superfluid component unpins and couples to the non-superfluid
component after time $\tg$, the superfluid transfers angular momentum
($\propto\Delta\nu_{\rm sf}$) to the rest of the star,
and a spin-up glitch $\Delta\nu$ is observed.
The pulsar then continues to spin down when the superfluid becomes
pinned again.}
\label{fig:glitch}
\end{figure}

\begin{figure}
\plotone{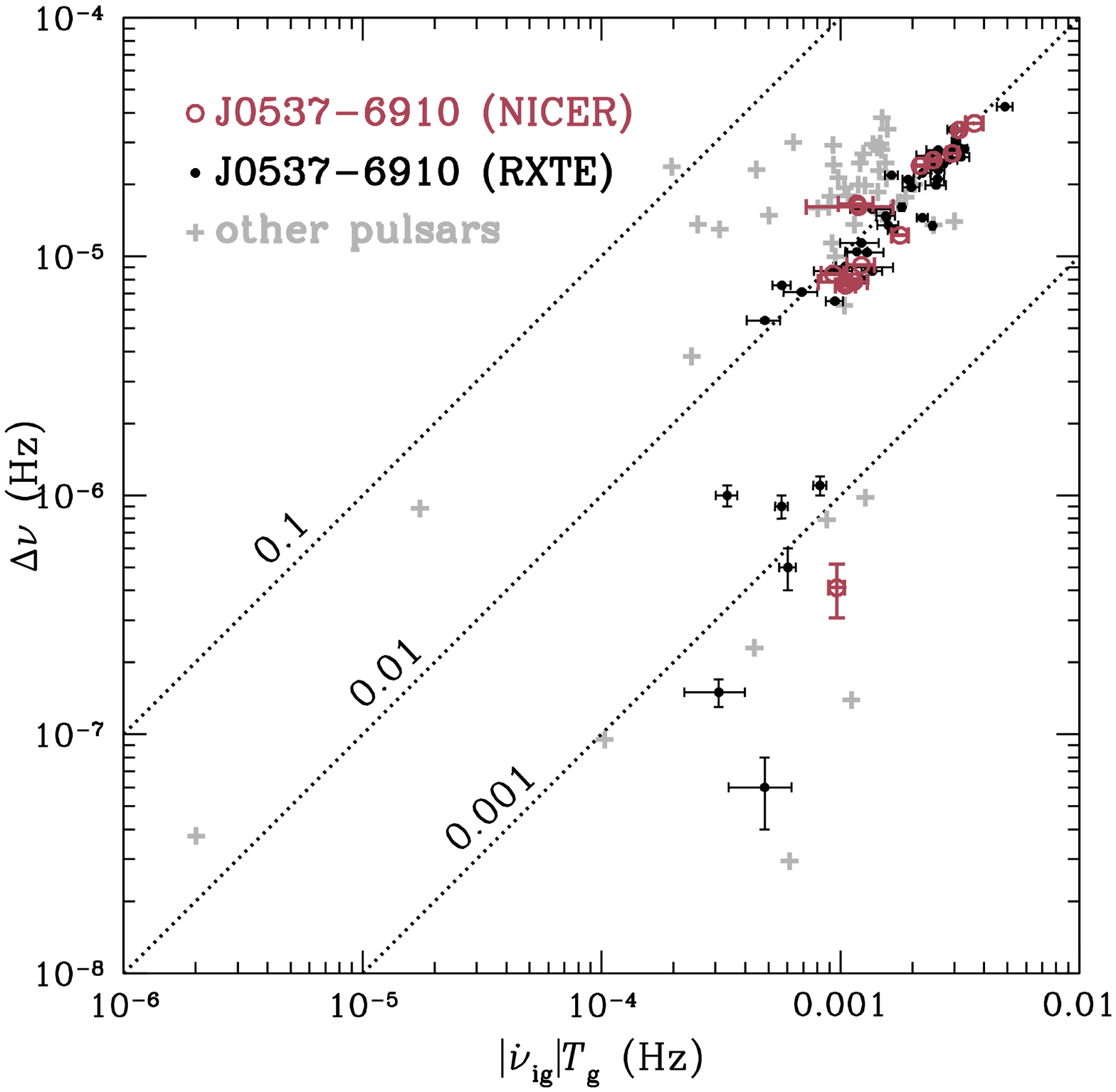}
\caption{Glitch size $\Delta\nu$, interglitch spin-down rate $\nudotig$,
and time to next glitch $\tg$ for \psrfive\ from NICER (large circles) and
RXTE (small circles), with the latter from \citet{antonopoulouetal18},
and for 16 other pulsars (crosses;
errors not shown since they are smaller than symbols in most cases)
from \citet{loweretal21}.
Dotted lines are $I_{\rm sf}/I=0.001,0.01,0.1$
[see equation~(\ref{eq:glitchcorr1}), where $\nudotig$ is
$\nudot$ between glitches and entrainment is neglected].
\label{fig:glitchcorr1}}
\end{figure}

One might expect pulsars with long observing times $\tobs$ that are
seen to glitch many times ($\Ng\gg1$, where $\Ng$ is number of
observed glitches) would show a correlation between the size of their
glitches $\Delta\nu$ and time between glitches
$\tg$,\footnote{It is often assumed that time between glitches $\tg$,
time since the previous glitch, and (wait) time to the next glitch
are all equivalent to each other.
However, these times are not necessarily the same in glitch models
(see, e.g., \citealt{carlinmelatos21}, and references therein) and do not
appear to be equivalent in observations.
For a few pulsars, there is a correlation between observed glitch sizes and
times to next glitch (see, e.g., Figure~\ref{fig:glitchpredict} for \psrfive).
But there is little support for a correlation between observed glitch
sizes and times to previous glitch in the case of \psrfive\
\citep{middleditchetal06,antonopoulouetal18,ferdmanetal18} and in
other glitching pulsars \citep{melatosetal18,fuentesetal19,loweretal21}.
Nevertheless, for the simple scenario outlined here, we assume these times
are equivalent; in particular, we use time to next glitch as $\tg$
for plotting glitch data in this section.}
which could be due to surpassing a critical threshold or building-up
or depleting an angular momentum reservoir.
Such a correlation can be derived by considering the simple
spin-down behavior of normal and superfluid components that is
illustrated in Figure~\ref{fig:glitch}.
A spin rate lag between these two components builds up during the time
between glitches $\tg$ due to a difference in spin-down rates of
the components, with $\nudot$ being the spin-down rate of the
normal component and $\nudot_{\rm sf}$ ($=0$) being that of the
superfluid.  At the glitch,
\begin{equation}
\Delta\nu+|\Delta\nu_{\rm sf}|=\nu_{\rm sf}-\nu =|\nudot|\tg, \label{eq:addspin}
\end{equation}
while angular momentum conservation implies
\begin{equation}
|\Delta\nu_{\rm sf}| = \frac{I}{I_{\rm sf}}\Delta\nu, \label{eq:angmom}
\end{equation}
where $I_{\rm sf}\ll I$ is assumed.
Combining equations~(\ref{eq:addspin}) and (\ref{eq:angmom}) yields
\begin{equation}
\Delta\nu = \frac{I_{\rm sf}}{I}|\nudot|\tg, \label{eq:glitchcorr1}
\end{equation}
which can also be related to glitch activity by taking $\tobs\approx\Ng\tg$
and $\sum\Delta\nu\sim\Ng\Delta\nu$ such that $\nudotg/|\nudot|=I_{\rm sf}/I$.
Equation~(\ref{eq:glitchcorr1}) can be used to constrain the
superfluid moment of inertia,
although in this case individual glitches are used rather than an
ensemble average which glitch activity parameters imply.
Figure~\ref{fig:glitchcorr1} illustrates this constraint
for $\nudotig$ equal to $\nudot$ measured between glitches
and for three values of $I_{\rm sf}/{I}$
(although recall we neglect entrainment, which would result in
a scaling factor of order unity in the constraint; see above),
as well as glitch data for \psrfive\ and some other pulsars.
Note that another superfluid constraint can be obtained by assuming
the spin-down torque on the entire star at a glitch is equal to the
torque on the normal (non-superfluid) component before or after the glitch,
which results in
\begin{equation}
\left|\frac{\Delta\nudot}{\nudot}\right| = \frac{I_{\rm sf}}{I}
\end{equation}
(see also \citealt{alparetal81}).

\begin{figure}
\plotone{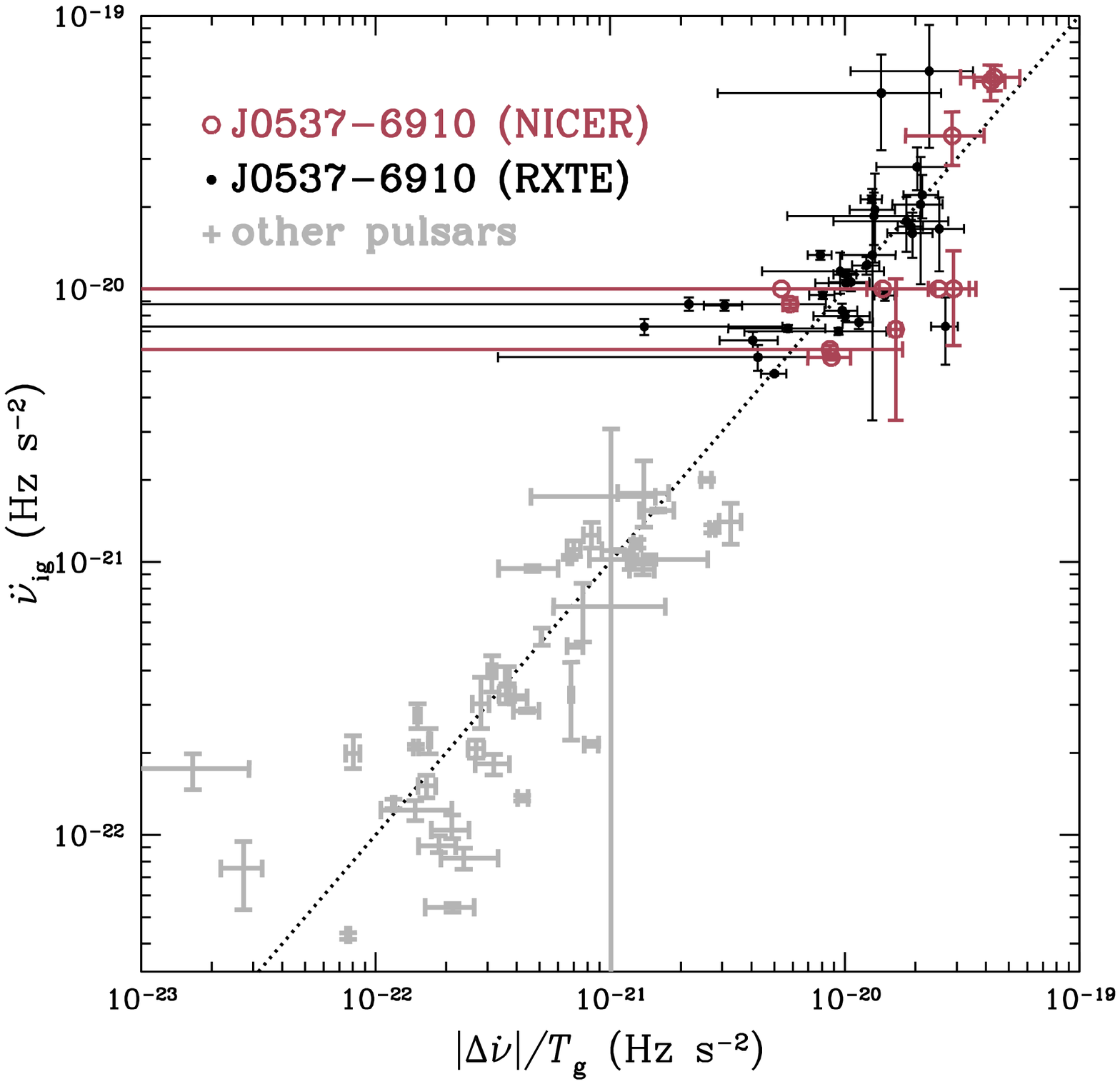}
\caption{Interglitch $\nuddotig$, glitch size $\Delta\nudot$,
and time to next glitch $\tg$ for \psrfive\ from NICER (large circles) and
RXTE (small circles), with the latter from \citet{antonopoulouetal18},
and for 16 other pulsars (crosses,
i.e., all data shown with $\nuddotig<3\times10^{-21}\mbox{ Hz s$^{-2}$}$)
from \citet{loweretal21}.
Dotted line is $\nuddotig=|\Delta\nudot|/\tg$
[see equation~(\ref{eq:glitchcorr2})].
\label{fig:glitchcorr2}}
\end{figure}

Another correlation can be obtained simply by assuming that
a glitch-induced change in spin-down rate $\Delta\nudot$ fully recovers
with time linearly at a rate $\nuddotig$, where $\nuddotig$ is
$\nuddot$ measured between glitches (see also \citealt{akbaletal17}).
This then implies
\begin{equation}
\left|\Delta\nudot\right| = \nuddotig\tg. \label{eq:glitchcorr2}
\end{equation}
Note that if the spin-down rate does not fully recover, then the
above can be replaced by $k|\Delta\nudot|=\nuddotig\tg$, where $k<1$.
Equation~(\ref{eq:glitchcorr2}) and glitch data from \citet{loweretal21}
and for \psrfive\ are plotted in Figure~\ref{fig:glitchcorr2}.
\citet{loweretal21} show that glitches of their 16 pulsars follow
the correlation, albeit with scatter.
We see here that the addition of glitches of \psrfive\ extends the
apparent correlation to much higher values of
$|\Delta\nudot|/\tg$ and $\nuddotig$.
Also note that we do not observe a strong correlation between $\nuddotig$
and $\delta t^2$, where $\delta t$ is time since previous glitch,
as expected from \citet{haskelletal20}.
However, we cannot make a firm conclusion since $\nuddotig$ is
likely strongly affected by glitch recovery effects in the first
$\sim$80 days after a glitch (see Figure~\ref{fig:nig}).
All the above and the issues discussed in Section~\ref{sec:0537}
illustrate the importance of continuing X-ray timing
observations and gravitational wave searches of \psrfive\ for its
impact on understanding glitches and revealing properties of dense
nuclear matter.

\begin{acknowledgments}

We thank S. Dzib for providing the radio position of \psreightone\ and
A. Kr\'olak, S. Mastrogiovanni, M. Pitkin, K. Riles, and G. Woan for support.
W.C.G.H. acknowledges support through grants 80NSSC21K0091 and
80NSSC21K1907 from NASA and Chandra award SAO GO1-22061X.
Chandra grants are issued by the Chandra X-ray Center (CXC),
which is operated by the Smithsonian Astrophysical Observatory
for and on behalf of NASA under contract NAS8-03060.
C.M.E. acknowledges support from ANID FONDECYT 1211964.
S.G. acknowledges the support of the Centre National d'Etudes Spatiales (CNES).
S.B. acknowledges support from NASA grant 80NSSC20K0275.
D.A. acknowledges support from an EPSRC/STFC fellowship (EP/T017325/1)
M.B. is partially supported by National Science Center (NCN) Poland
grants 2016/22/E/ST9/00037 and 2017/26/M/ST9/00978.
B.H. acknowledges support from NCN Poland via SONATA BIS grant
2015/18/E/ST9/00577.

This research made use of data obtained from the Chandra Data
Archive and the Chandra Source Catalog, and software provided by the
Chandra X-ray Center (CXC) in the application packages CIAO and Sherpa.
This work is supported by NASA through the NICER mission and the
Astrophysics Explorers Program and uses data and software provided
by the High Energy Astrophysics Science Archive Research Center
(HEASARC), which is a service of the Astrophysics Science Division
at NASA/GSFC and High Energy Astrophysics Division of the
Smithsonian Astrophysical Observatory.
This work made extensive use of the NASA Astrophysics Data System
(ADS) Bibliographic Services and the arXiv.

The Fermi LAT Collaboration acknowledges generous ongoing support
from a number of agencies and institutes that have supported both the
development and the operation of the LAT as well as scientific data analysis.
These include NASA and 
the Department of Energy (DOE) in the United States,
the Commissariat \`a l'Energie Atomique and the Centre National de la
Recherche Scientifique/Institut National de Physique Nucl\'eaire et de
Physique des Particules in France,
the Agenzia Spaziale Italiana and the Istituto Nazionale di Fisica Nucleare
in Italy,
the Ministry of Education, Culture, Sports, Science and Technology (MEXT),
High Energy Accelerator Research Organization (KEK) and Japan Aerospace
Exploration Agency (JAXA) in Japan, and
the K.~A.~Wallenberg Foundation, the Swedish Research Council and the
Swedish National Space Board in Sweden.
Additional support for science analysis during the operations phase is
gratefully acknowledged from the Istituto Nazionale di Astrofisica in
Italy and the Centre National d'\'Etudes Spatiales in France.
This work performed in part under DOE Contract DE-AC02-76SF00515.
\end{acknowledgments}

\vspace{5mm}
\facilities{Chandra, Fermi, NICER, NuSTAR}

\software{CIAO (\url{https://cxc.harvard.edu/ciao/}),
HEAsoft (\url{https://heasarc.nasa.gov/lheasoft/}),
PINT (\url{https://github.com/nanograv/pint}),
PRESTO (\url{https://www.cv.nrao.edu/~sransom/presto/}),
TEMPO2 \citep{hobbsetal06}}

\bibliography{arxiv}{}
\bibliographystyle{aasjournal}

\end{document}